\documentclass[%
 reprint,
 aps, ]{revtex4-2}  \usepackage{graphicx, float}
\usepackage{dcolumn}
\usepackage{bm, braket, bbold}
\usepackage{hyperref}
\usepackage{mathtools}
\usepackage{amsmath,amssymb}
\hypersetup{
    colorlinks=true
}
\usepackage{todonotes}
\usepackage{appendix}
\usepackage[acronym]{glossaries}
\newacronym{mp}{MP}{Mutual Predictability}
\newacronym{mi}{MI}{Mutual Information}
\newacronym{pcc}{PCC}{Pearson Correlation Coefficient}
\newacronym{npt}{NPT}{Negative Partial Transpose}

\newcommand{\abb}[1]{{\hypersetup{linkcolor=black}\gls{#1}}}
\newcommand{\E}{\mathrm{e}}
\newcommand{\I}{\mathrm{i}}

\begin{document}

\preprint{APS/123-QED}

\title{Relating an entanglement measure with statistical correlators for two-qudit mixed states using only a pair of complementary observables}

\author{Simanraj Sadana$^{1,\dagger}$, Som Kanjilal$^{2,\ddagger}$, Dipankar Home$^{2}$, Urbasi Sinha$^{1}$}
\email{usinha@rri.res.in}
\affiliation{$^{1}$Light and Matter Physics, Raman Research Institute, Bengaluru-560080, India}
\affiliation{$^{2}$Center for Astroparticle Physics and Space Science (CAPSS),Bose Institute, Kolkata 700 091, India.}
\altaffiliation{Now at \emph{Istituto Nazionale di Fisica Nucleare, Sezione di Pavia, Via Agostino Bassi 6, 27100, Pavia, PV, Italy.}\\
${}^\ddagger$Now at \emph{Harish-Chandra Research Institute Chhatnag Road, Jhunsi, Prayagraj (Allahabad) 211 019 India.}}{}

\begin{abstract}
We focus on characterizing entanglement of high dimensional bipartite states using various statistical correlators for two-qudit mixed states. The salient results obtained are as follows: (a) A scheme for determining the entanglement measure given by Negativity is explored by analytically relating it to the widely used statistical correlators viz. mutual predictability, mutual information and Pearson Correlation coefficient for different types of bipartite arbitrary dimensional mixed states. Importantly, this is demonstrated using only a pair of complementary observables pertaining to the mutually unbiased bases. (b) The relations thus derived provide the separability bounds for detecting entanglement obtained for a fixed choice of the complementary observables, while the bounds per se are state-dependent.  Such bounds are compared with the earlier suggested separability bounds. (c) We also show how these statistical correlators can enable distinguishing between the separable, distillable and bound entanglement domains of the one-parameter Horodecki two-qutrit states. Further, the relations linking Negativity with the statistical correlators have been derived for such Horodecki states in the domain of distillable entanglement. Thus, this entanglement characterisation scheme based on statistical correlators and harnessing complementarity of the obsevables opens up a potentially rich direction of study which is  applicable for both distillable and bound entangled states.

\end{abstract}
\maketitle
The characterisation of quantum entanglement is of key importance while using it as a resource for its multifarious applications in the frontier areas of quantum technology like quantum communications, quantum computation, and quantum metrology. While the workhorse of the related implementations to date has been primarily the entanglement between qubits, it is now well recognised that the increased dimensionality of the entangled states can contribute significantly in enhancing the efficiency and robustness of their use as a resource \cite{Bennett1999,Bechman2000,Cerf2002,Vertesi2010,Sheridan2010,Bruss2003,Wang2005}. Hence, the issue of characterising any arbitrary high-dimensional entangled state has gained considerable significance, wherein the characterisation entails both the device-independent certification and the quantification in terms of the appropriate entanglement measures (EMs) defined as entanglement monotones which are non increasing under LOCC and vanishing for separable states. Here we should note that for any such higher dimensional entanglement characterisation scheme to be operationally effective, it needs to be based on a limited number of measurements for any state dimension, in contrast to any method using the tomographic technique for characterising a quantum state that requires a rapidly increasingly large number of measurements as the dimension of the state space increases.\cite{huang_2014,harrow2017}

In this work, we focus on characterizing bipartite arbitrary dimensional entangled states based on measurements of complementary observables using at most a pair of mutually unbiased bases (MUBs), the correlations therein being captured in terms of the standard statistical correlators like \abb{mp}, \abb{mi} and \abb{pcc}. For this purpose, on the one hand, we consider a range of distillable entangled bipartite states with noises like the isotropic noise, coloured noise A, coloured noise B, and Werner states for which we are able to analytically relate the statistical correlators MP, MI and PCC to the measure of entanglement given by Negativity, valid for any dimension. On the other hand, we also consider the non-distillable entangled state, known as the bound entangled state, such as the one-parameter Horodecki states \cite{Horodecki1999}. For such states, we show that \abb{mp}, \abb{mi} and \abb{pcc} can be used to certify entanglement in the non-distillable regime of these states, apart from analytically relating \abb{mp}, \abb{mi} and \abb{pcc} to Negativity in the distillable (\abb{npt}) regime of such states. Thus, the results obtained in this study provide an emphatic illustration of the efficacy of the use of statistical correlators and mutually unbiased bases for characterizing high dimensional entanglement. \\

The choice of distillable entangled states for  this study has been particularly motivated by the empirical consideration, focusing on the noisy Bell-state which is a convex combination of the above mentioned usually prevalent noises. We show that, for a particular pair of complementary measurement bases, the Negativity of such a state can be expressed as a function of the measured quantity \abb{mp} and the knowledge of the total amount of noise, irrespective of the amounts of individual noises. We also show that \abb{mi} and \abb{pcc} have similar relations with the Negativity of noisy Bell-states. Now, an important point is that for a given choice of the complementary measurement bases, the relation between the entanglement measure and the statistical correlators depends on the type of state under consideration. In order to highlight this feature, we find such relations for a couple of other classes of states. For example, in the case of the Werner state \cite{Werner1989}, such a relation is different from that for the noisy Bell-state. Therefore, it is necessary to have knowledge of the type of state in order to use such relations. Nevertheless, this method is useful in experiments in which a source is designed to produce a particular state, but due to the presence of noise, the fidelity is not guaranteed to be unity. If the type of noise relevant to a given experimental context is known or can be reasonably guessed, then the state-dependent relations mentioned earlier can be used to determine the Negativity by measuring the statistical correlators. The dependence of these relations on the type of state is the result of using a fixed set of complementary observables for any state. This is in contrast with the state-independent separability criteria derived/conjectured in \cite{Spengler2012, Maccone2015}, but which depend on choosing the specific measurement bases which are contingent on the given state. The comparison between these two approaches is analyzed in the present paper, alongside proving a relevant conjecture \cite{Maccone2015} for pure and colored noise A states.\\

Another salient feature of this paper is that, we demonstrate the utility of the statistical correlators for detecting non-distillable entanglement, using the example of the one-parameter Horodecki (OPH) states. First, we show that MP, MI and PCC can be used to distinguish between separable, distillable and non-distillable entangled OPH states, enabling to define an ordered relation over the set of OPH states, based on the value of the statistical correlator. The interpretation and the use of such an ordered relationship motivates an interesting line of study. Secondly, for the distillable OPH states, the monotonic relations between the Negativity and statistical correlators have been obtained.\\

The paper is organised as follows. In \S\ref{sec:history} we provide the relevant background of the overall studies towards entanglement characterization. In \S\ref{sec:main_results} we present our main results for the determination of Negativity in terms of the statistical correlators for the classes of states we have considered. In \S\ref{sec:separability_bounds} we discuss the implications of our work pertaining to the earlier proposed separability bounds by Spengler and Maccone et al.~\cite{Spengler2012, Maccone2015}, and compare them with the separability bounds obtained in our treatment. In this process, we also provide analytical justification of the separability bound proposed by Maccone et. al.~\cite{Maccone2015} for certain states. In the concluding section \S\ref{sec:summary}, the summary of our work is outlined and the indications are given of a few possible future directions of studies.

\section{Background}\label{sec:history}
The simplest device-independent scheme for certifying entanglement, such as the one using the violation of Bell inequality does not provide the necessary and sufficient condition for detecting entanglement, since there are entangled states which do not violate the Bell-inequality. The other basic scheme invoking the Partial-Positive-Transpose(PPT) criterion for separability is also of limited use, since it provides the necessary and sufficient condition for detecting entanglement essentially restricted to 2x2 and 2x3 systems. In fact, the question of device-independent characterisation of a wide class of entangled states in any arbitrary dimension is, in general, a formidable open problem.

A widely discussed approach to address this issue is known as the Entanglement Witness (EW) based approach \cite{Terhal2000}. The Hahn-Banach theorem \cite{Edwards2012} guarantees the existence of an observable (called an EW) whose measured expectation value provides a necessary and sufficient condition for distinguishing a given entangled state from all the separable states. However, finding a suitable witness that satisfies this condition, without any knowledge of the state in hand is not easy. This is why the problem of an EW being sufficient but not necessary often arises. The negative expectation values of EWs have been used to put a lower-bound on certain measures of entanglement, by optimising the relevant entanglement measure subject to the limited (measured) data that is available. There are approaches in which the lower-bound is estimated using the diagonal elements and only a few off-diagonal elements of the density matrix, thereby reducing the required number of measurements significantly \cite{Tiranov2017, Martin2017}. However, conventional EW measurements suffer from measurement based imperfections. One of the goals of entanglement-certification protocols is thus to be device-independent as imperfect measurements can lead to erroneous construction of EW and possibly an incorrect conclusion about the presence of entanglement. To account for this, measurement device independent entanglement witness (MDIEW) schemes have been developed to detect entanglement even with imperfect measurements. For example, one such protocol builds on the use of non-local quantum games to verify entanglement \cite{Buscemi2012}, and the other one uses standard EWs to construct a witness for entanglement which is robust against imperfect measuring devices \cite{Branciard2013}. A recent work has resulted in a ``quantitative" MDIEW scheme \cite{Guo2020}. However, this work is limited in its scope as it provides a lower bound on an entanglement measure, which is not a conventional EM. In particular, this scheme is essentially applicable to a class of entangled states that the authors have called ``irreducible entanglement"\cite{Guo2020}.

Now, we proceed to discuss schemes that have been developed independent of the EW based line of studies. Approaches have been proposed for quantifying entanglement by determining the Schmidt number of a mixed entangled state in any dimension \cite{Shahandeh2013, Shahandeh2014}. For any pure state, the Schmidt number is the same as the Schmidt rank. For a mixed state, the Schmidt number denotes the maximal local dimension of any pure state contribution to the mixed state. It has been argued that the Schmidt number satisfies the requirements of an entanglement monotone \cite{sperling2011schmidt}. However, the number of measurements required to determine the Schmidt number scales up with the dimension of the state, and hence it is difficult to implement as the dimension increases. 

Next, coming to the approaches seeking to quantify entanglement on the basis of a limited number of measurements, all of them quantify entanglement in terms of the lower bounds of EMs. For example, the scheme proposed by \cite{bavarescoJ} can determine the lower bounds of both the Schmidt number as well as of EOF by relating these respective bounds to a suitable quantifier of the mixedness of the prepared state, based on only two local measurement bases in each wing, which are not required to be mutually unbiased. 

On the other hand, there are schemes using two mutually unbiased bases that enable determining the lower bound of EOF by relating it to respectively the quantum violation of the EPR-steering inequality \cite{Schneeloch2018} and a combination of suitably defined correlators \cite{Erker2017} which are not the standard statistical correlators like Mutual Predictability (MP), Mutual Information (MI) and Pearson Correlation Coefficient (PCC).

The use of the standard statistical correlators like MP, MI and PCC for the purpose of characterising bipartite arbitrary dimensional entanglement, by employing two pairs of mutually unbiased bases, was invoked by \cite{Spengler2012}, followed by \cite{Maccone2015} . These studies have essentially focused on deriving the separability bounds using such statistical correlators for certifying arbitrary dimensional entanglement of pure and some specific mixed two-qudit states.

A key shift in focus from the above-mentioned line of studies has been made in the latest study reported in a separate paper \cite{g+22}, where by harnessing the maximum complementarity of one or two pairs of mutually unbiased measurement bases, the EMs like Negativity (N) and Entanglement of Formation (EOF) of any bipartite pure state have been analytically related to the observable statistical correlators like MP, MI and PCC for any dimension. Moreover, this general scheme has been applied for the experimental determination of N and EOF of a prepared two qutrit pure photonic state, using the analytically derived relationships. It is this newly launched line of study which is comprehensively developed in the present paper for a range of bipartite arbitrary dimensional mixed states - isotropic, coloured noise A, coloured noise B, and Werner states.

At this stage, a couple of points need to be noted:
a) An earlier preliminary study had investigated the use of PCCs for determining N of a class of bipartite entangled states, but this study was restricted to dimensions 3, 4 and 5 \cite{Jebarathinam2020}.
b) A scheme was earlier proposed to relate the EMs like the Concurrence and Negativity of pure and isotropic mixed states to the values of the Bell-SLK function evaluated using a limited number of observables \cite{Datta2017}. However, the experimental realisability of this scheme remains unexplored, and the study has not yet been extended to other types of mixed states.

\section{Determination of Negativity using statistical correlators}\label{sec:main_results}
The general approach here is the following. Given a parametric bipartite quantum state $\rho(\{\alpha_i\})$, let the Negativity of the state be $\mathcal{N}(\{\alpha_i\}$ defined as,
\begin{align}
    \mathcal{N} = \frac{\|{\rho^{T_A}}\| - 1}{2}
\end{align}
and $C(\{\alpha_i\})$ be the statistical correlator of choice. The relation between the Negativity and the statistical correlator is then found by solving the parametric equations.

The statistical correlators are calculated for measurements in two complementary measurement bases. We choose the eigenbases of the generalized Pauli operators $X$ and $Z$ for such measurements. The three statistical correlators that we consider are MP, MI and PCC.

MP of a $d\times d$ bipartite state $\rho$ with respect to the two chosen bases $\{\ket{a_i, b_j}\}$ is defined as
\begin{align}
    P_{AB} = \sum\limits_{i=0}^{d} \braket{a_i,b_i |\rho|a_i,b_i}
\end{align}
Here for all the cases considered, we fix the measurement bases to be the eigenbases of two complementary observables $Z$ and $X$ defined as
\begin{align}
    Z\ket{i} =& \omega_d^i \ket{i} \\
    X\ket{i} =&\ket{i \oplus_d 1}
\end{align}
where $\omega_d = \E^{2 \pi \I/d}$.
MP with respect to the eigenbasis of $Z\otimes Z$ (or $X \otimes X$) will henceforth be denoted by $P_Z$ (or $P_X$).

MI is defined as
\begin{align}
    I_{AB} =& \sum\limits_{i,j = 0}^{d,d} p(i,j) \log_2\left(\frac{p(i,j)}{p_a(i) p_b(j)}\right)
\end{align}
where
\begin{align}
    p(i,j) =& \braket{a_i,b_j | \rho | a_i, b_j} \\
    p_a(i) =& \braket{a_i | \mathrm{Tr}_b \rho| a_i}
\end{align}
and PCC is defined as
{\small
\begin{align}
\label{eq:exprpcc}
    PCC_{AB} =& \frac{\sum\limits_{i,j} p(i,j) x_i x_j - \sum\limits_{i} p_a(i) x_i \sum\limits_{j} p_b(j) x_j}{\sqrt{\sum\limits_{i} p_a(i) x_i^2 - \left(\sum\limits_{i} p_a(i) x_i\right)^2}\sqrt{\sum\limits_{j} p_b(j) x_j^2 - \left(\sum\limits_{j} p_b(j) x_j\right)^2}}
\end{align}
}
\subsection{Noisy Bell state}\label{sec:noisy_bells}
Here we consider Bell states with three types of noises, based on their prevalence in the experiments, wiz., isotropic/white noise, coloured noise A and coloured noise B defined respectively as
\begin{align}
    \label{eq:isotropic_noise}
    \rho_\mathrm{iso} =& \frac{\mathbb{1}}{d^2}, \\
    \label{eq:colourA}
    \rho_\mathrm{cna} =& \frac{1}{d} \sum\limits_{i=0}^{d-1} \ket{i,i}\bra{i,i}, \\
    \label{eq:colourB}
    \rho_\mathrm{cnb} =& \frac{1}{d(d-1)} \sum\limits_{\substack{i,j=0\\i\neq j}} \ket{i,j} \bra{i,j},
\end{align}
for a $d \times d$ system. Isotropic/white noise is the simplest yet most prevalent type of noise, which probabilistically converts a qudit into a maximally mixed state. For example, when a polarized photonic qubit passes through a depolarising channel (e.g. atmosphere in free-space quantum communication), it incurs white noise as the polarisation of some photons is completely randomised \cite{Keyl2002}. Due to the $U \otimes U^*$ symmetry of the isotropic noise, it is often used a test-bed for studying entanglement \cite{Tsokeng2018, Derkacz2006}. On the other hand, coloured noise A has a special property of being completely correlated in the computational basis whereas coloured noise B is completely anti-correlated \cite{Jebarathinam2020}. The former is produced in the Hong-Ou-Mandel interferometers using polarisation flipper \cite{Huang2016}, and in the case of $d=2$, coloured noise B occurs when one of the parties (of the bipartite system) goes through a local dephasing channel \cite{xiang}. Coloured noise B is also important because it has symmetries equivalent to those of a maximally entangled state \cite{Sentis2016QuantifyStates}.

We consider a convex combination of a Bell state and the three types of the above mentioned noises 
\begin{align}
    \label{eq:noisy_Bell}
    \rho =& a \ket{\phi^+}\bra{\phi^+} + (1 - a) \left(b \rho_\mathrm{iso} + (1-b) \left(c\rho_\mathrm{cna}\right.\right.\nonumber\\ 
    &\left.+ (1-c) \rho_\mathrm{cnb}\right)
\end{align}
where the range of the parameters is $0 \leq a,b,c \leq 1$ and 
\begin{align}
    \label{eq:phi_plus}
    \ket{\phi^+} =& \frac{1}{\sqrt{d}} \sum\limits_{i=0}^{d-1} \ket{i,i}
\end{align}
is a maximally entangled Bell state. The results for the other Bell states, viz., $\ket{\phi^-}$, $\ket{\psi^+}$ and $\ket{\psi^-}$ are similar.

The Negativity of the noisy Bell state given by Eq. \eqref{eq:noisy_Bell} is 
\begin{align}
    \label{eq:neg_noisy_bell}
    \mathcal{N} =& \max\left\{ 0, \frac{1}{2d}\left[-d + ad^2 + b -ab + d (1-a)(1-b)c\right] \right\}
\end{align}
where entangled state is distillable for those values of $a,~b,~c,~d$ for which $\mathcal{N} \neq 0$. Note that, by applying suitable variable transformation, the noisy Bell state can be shown  equivalent to the class of states studied in \cite{Eltschka2013}. The probabilities of outcomes of joint local measurements performed on this state with respect to the $Z$ and $X$ bases are respectively (see Appendix \ref{app:noisy-bell-probs} for details)
\begin{align}
    \label{eq:joint_prob_noisy_bell}
    p_Z(i,j) =& a \frac{\delta_{ij}}{d} + \left(1 - a\right) \left(\frac{b}{d^2} \right.\nonumber\\ 
    &+ \left. \left(1-b\right) \left(c \frac{\delta_{ij}}{d} + \left(1-c\right) \frac{1-\delta_{ij}}{d(d-1)} \right)\right) \\
    \label{eq:joint_prob_noisy_bell2} 
    =& \begin{cases}\frac{1-a(d-1)+2\mathcal{N}}{d} & i=j \\
    \frac{a}{d} - \frac{2\mathcal{N}}{d(d-1)} & i\neq j\end{cases} \\
    p_X(i,j) =& a \frac{\delta_{ij}}{d} + (1-a) \frac{1}{d^2}
\end{align}
Note that the quantities $p_z(i,i)$ and $p_z(i,j|i\neq j)$ are functions of $a$ and $\mathcal{N}$ but are independent of $b$ and $c$. Therefore, all the three statistical correlators MP, MI and PCC with respect to the $Z$ basis must be functions of only $a$ and $\mathcal{N}$. This implies that $\mathcal{N}$ can be expressed as a function of the statistical correlators and $a$, where the value of $a$ can be found from the values of these statistical correlators with respect to the $X$ basis. We plot these relations for the different correlators and it is seen that $\mathcal{N}$ for the noisy Bell-states can be calculated from these statistical correlators.

\begin{figure}
    \centering
    \includegraphics[width=1\hsize]{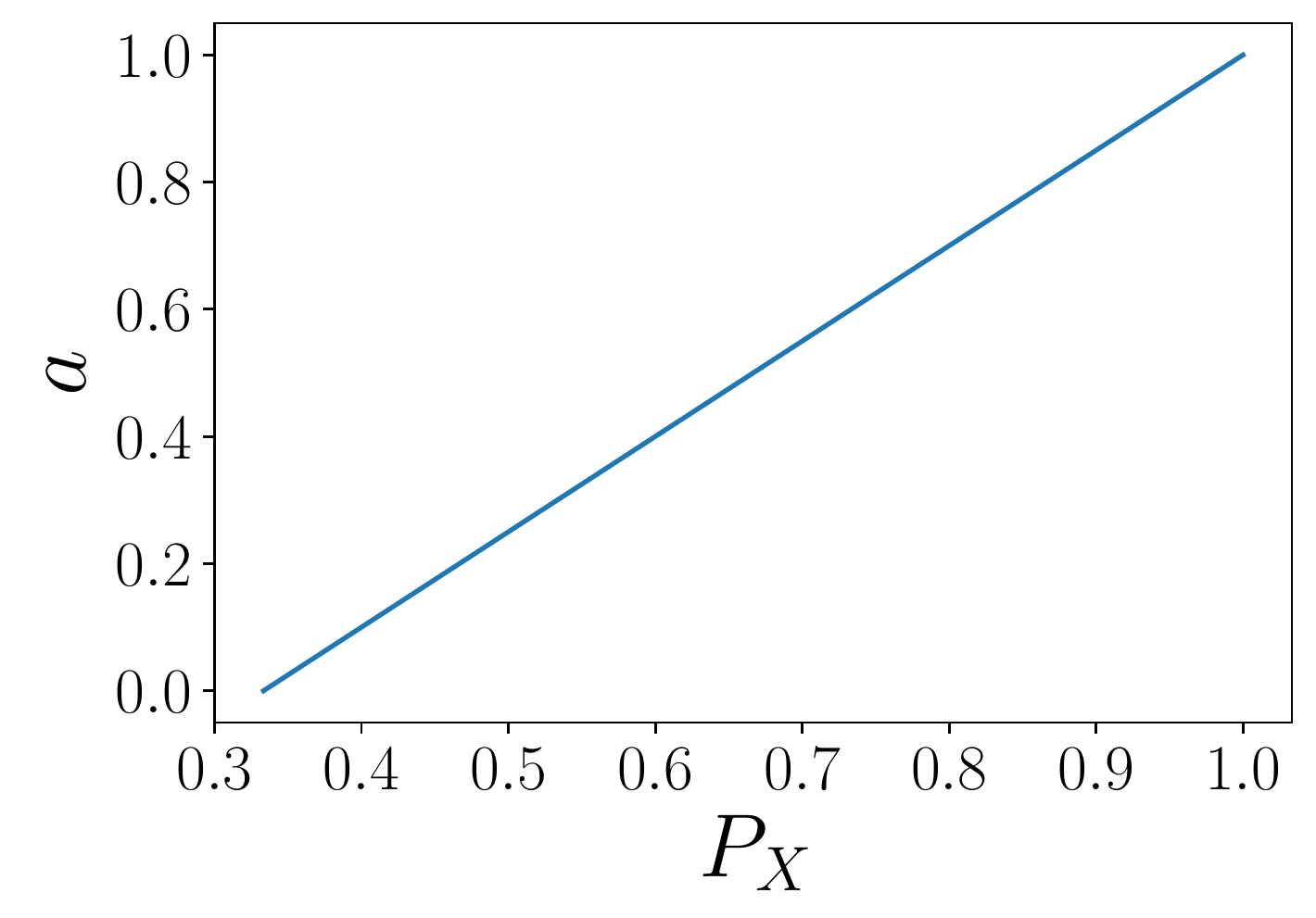}
    \includegraphics[width=1\hsize]{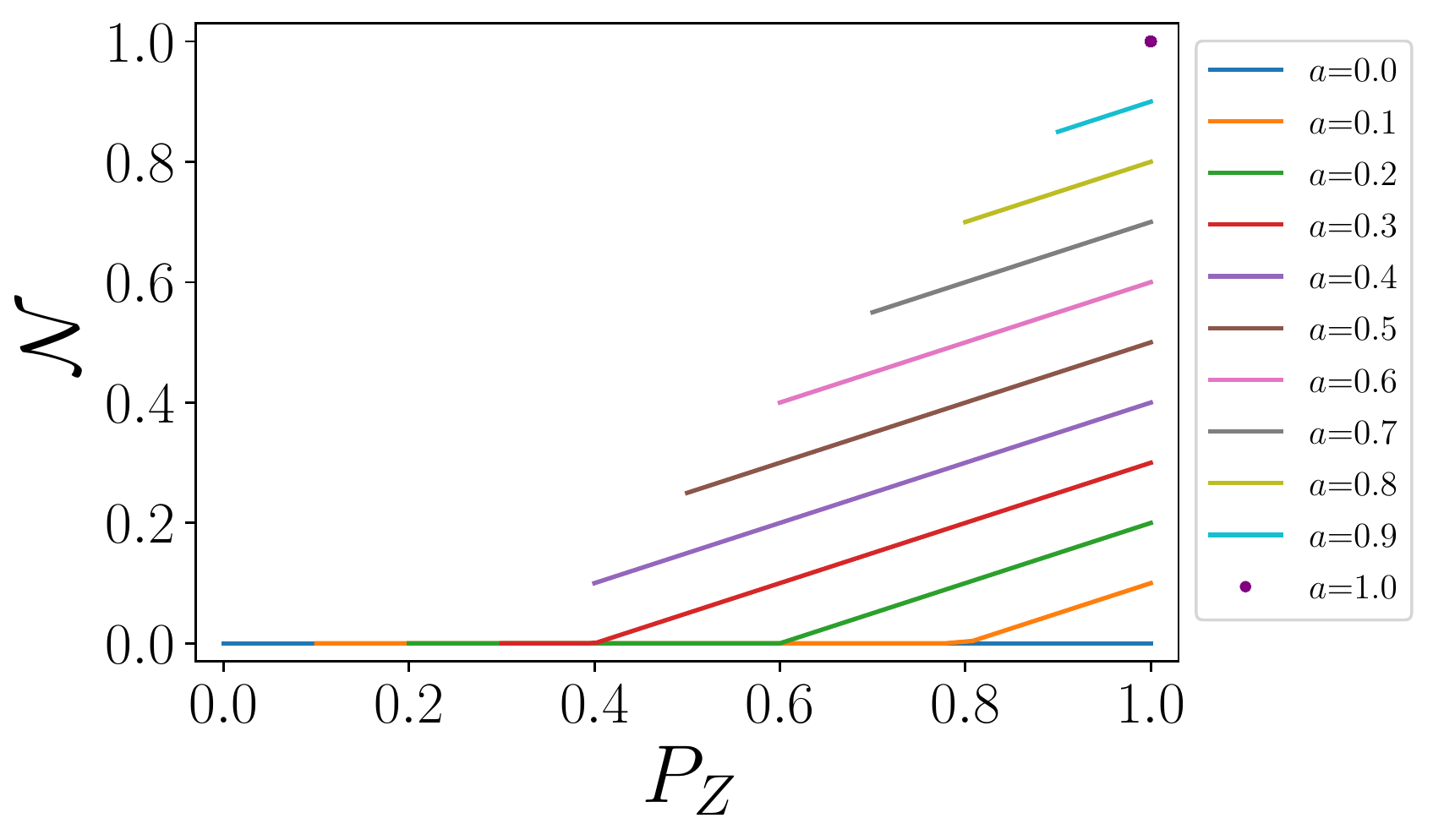}
    \caption{(Top) The plot of the parameter $a$ versus $P_X$ (MP with respect to the $X$ basis), for the noisy Bell-state of Eq.~\eqref{eq:noisy_Bell} with $d=3$. (Bottom) The plot of Negativity ($\mathcal{N}$) of the noisy Bell state versus $P_{Z}$ (MP with respect to the $Z$ basis), for different values of $a$. The value of $a$ is found from the value of $P_X$ from the top graph, and then using this value of $a$, the bottom graph is used to find $\mathcal{N}$ from the value of $P_Z$.}
    \label{fig:mpxmpz_noisy_bell}
\end{figure}

\begin{figure}
    \centering
    \includegraphics[width=1\hsize]{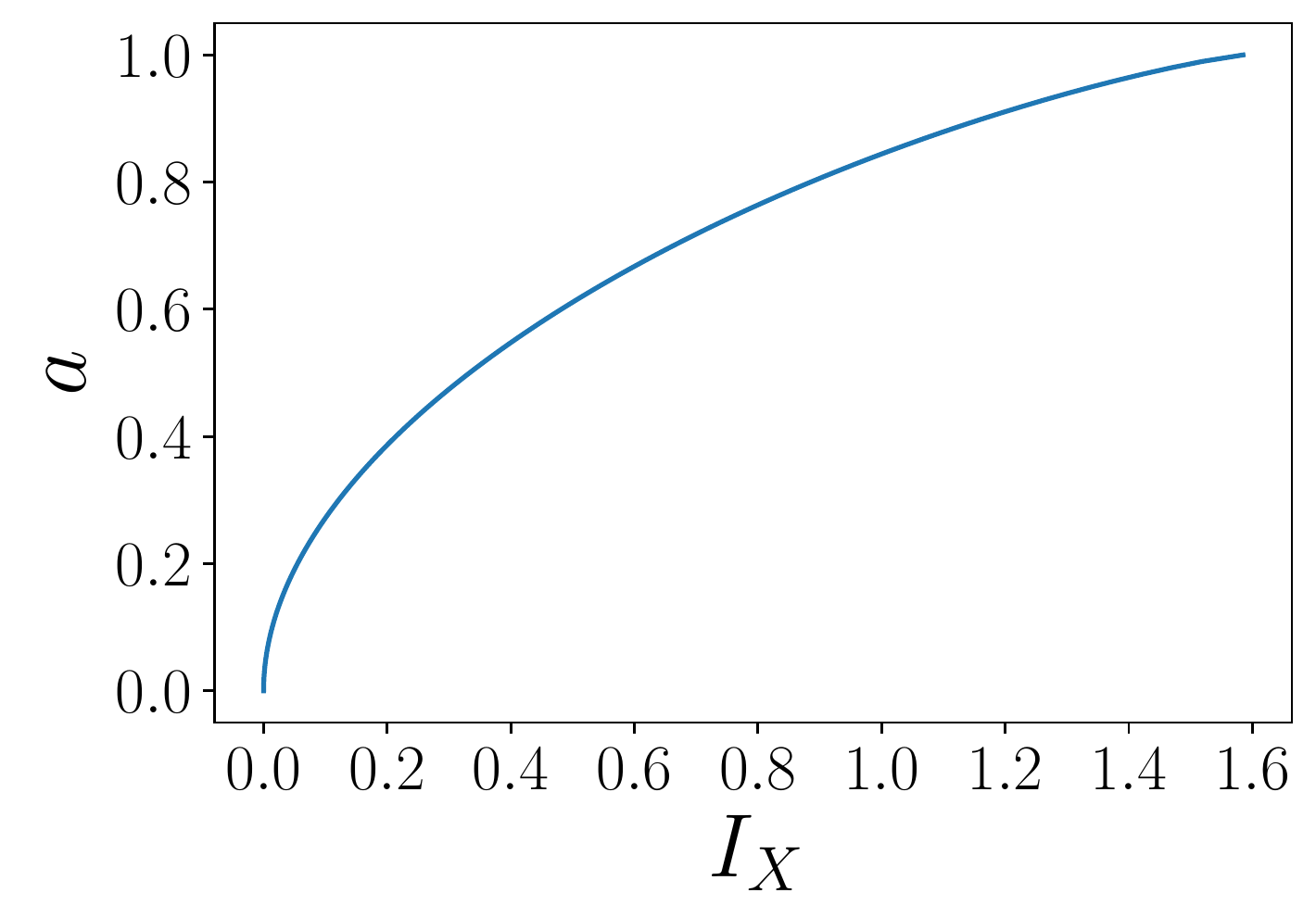}
    \includegraphics[width=1\hsize]{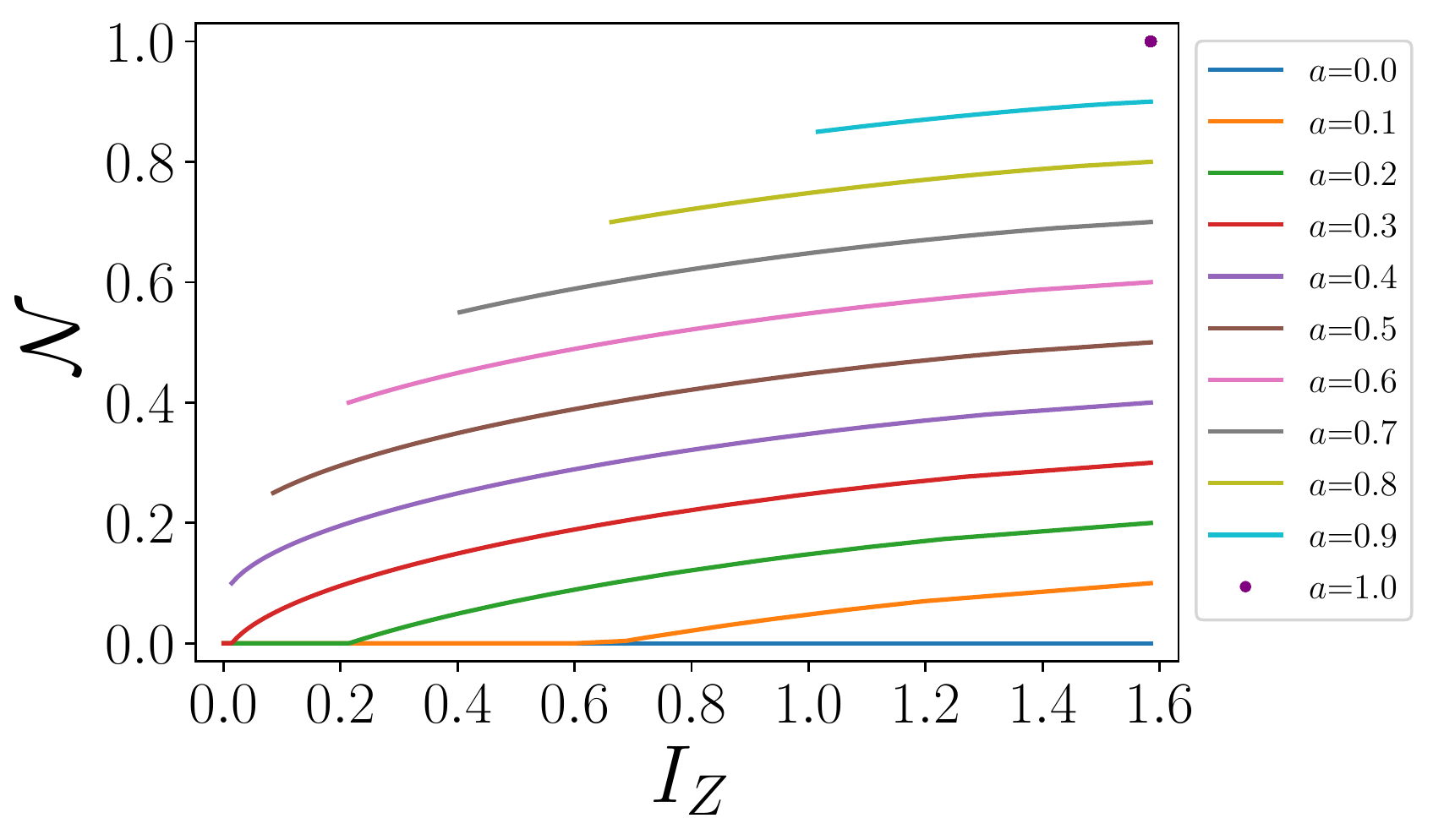}
    \caption{(Top) The plot of the parameter $a$ versus $I_X$ (MI with respect to the $X$ basis), for the noisy Bell state of Eq.~\eqref{eq:noisy_Bell} with $d=3$. (Bottom) The plot of Negativity ($\mathcal{N}$) of the noisy Bell state versus $I_Z$ (MI with respect to the $Z$ basis), for different values of $a$. The value of $a$ is found from the value of $I_X$ from the top graph, and then using this value of $a$, the bottom graph is used to find $\mathcal{N}$ from the value of $I_Z$.}
    \label{fig:mi_combined_noises}
\end{figure}

\begin{figure}
    \centering
    \includegraphics[width=1\hsize]{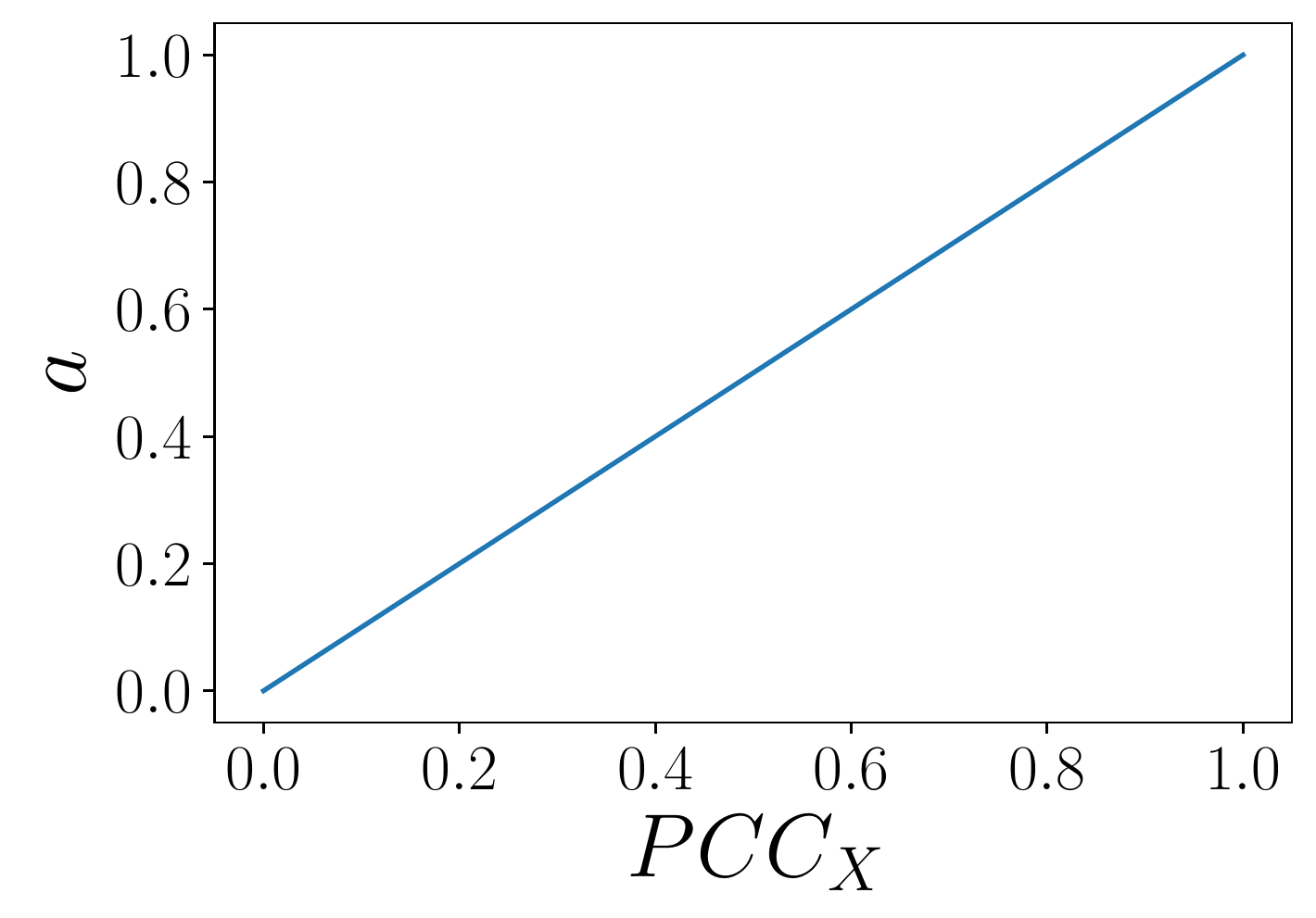}
    \includegraphics[width=1\hsize]{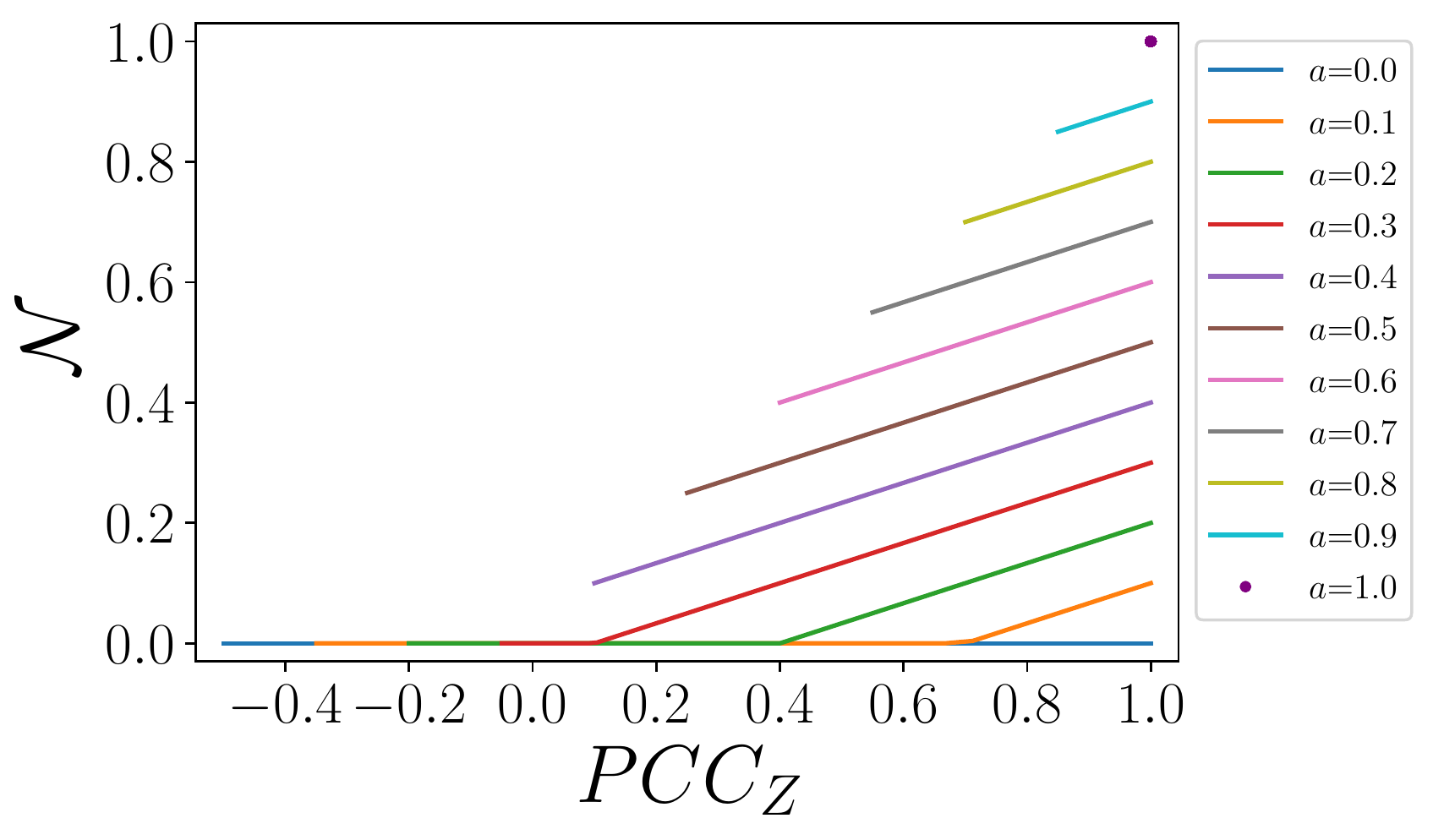}
    \caption{(Top) The plot of the parameter $a$ versus $PCC_X$ (PCC with respect to the $X$ basis), for the noisy Bell state of Eq.~\eqref{eq:noisy_Bell} with $d=3$. (Bottom) The plot of Negativity ($\mathcal{N}$) of the noisy Bell state versus $PCC_Z$ (PCC with respect to the $Z$ basis), for different values of $a$. The value of $a$ is found from the value of $PCC_X$ from the top graph, and then using this value of $a$ the bottom graph is used to find $\mathcal{N}$ from the value of $PCC_Z$.}
    \label{fig:pcc_noisy_bell_state}
\end{figure}

Figs. \ref{fig:mpxmpz_noisy_bell}, \ref{fig:mi_combined_noises} and \ref{fig:pcc_noisy_bell_state} respectively shows the plots for MP, MI and PCC which indicate the way these statistical correlators can be used to determine $\mathcal{N}$ for the noisy Bell states for $d=3$. An interesting feature of these results is that the amount of an individual type of noise present in the state is irrelevant and $\mathcal{N}$ depends only on the total amount of noise, given by $a$. 

As mentioned earlier, the monotonic relations between $\mathcal{N}$ and the statistical correlators are dependent on the type of state if the measurement bases are fixed. To illustrate this, we now show different such relation for a different state.


\subsection{Werner state}\label{sec:werner}
Werner state is an important quantum state, that provided the first example of a mixed-entangled state which does not violate Bell-inequality and admits a local realist model. It is a $U \otimes U$ invariant state that has found uses in the studies of quantum entanglement like those concerning the non-additivity of bipartite distillable entanglement, deterministic purification of noisy entangled state and transport of entanglement over noisy channels.

\begin{figure}[h]
\centering
\includegraphics[width=1\hsize]{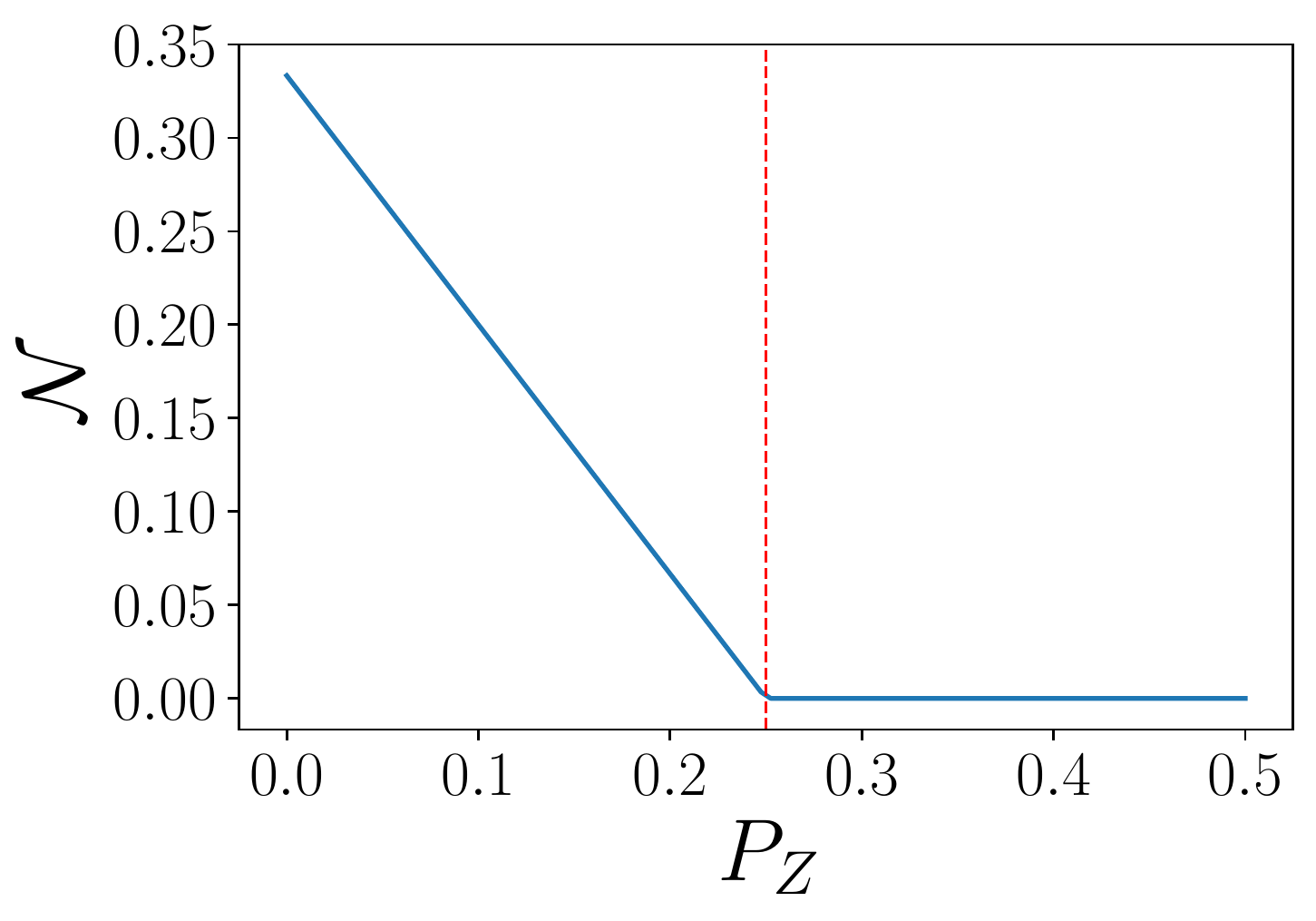}
\caption{Negativity $(\mathcal{N})$ of the Werner state as a function of MP ($P_{z}$) with respect to the $Z$-basis. The plot shows that for the entangled Werner states, $P_Z$ has an upper bound. Below that bound, $\mathcal{N}$ depends linearly on the value of MP.}
\label{fig:mp_werner_comp}
\end{figure}

\begin{figure}[h]
    \centering
    \includegraphics[width=1\hsize]{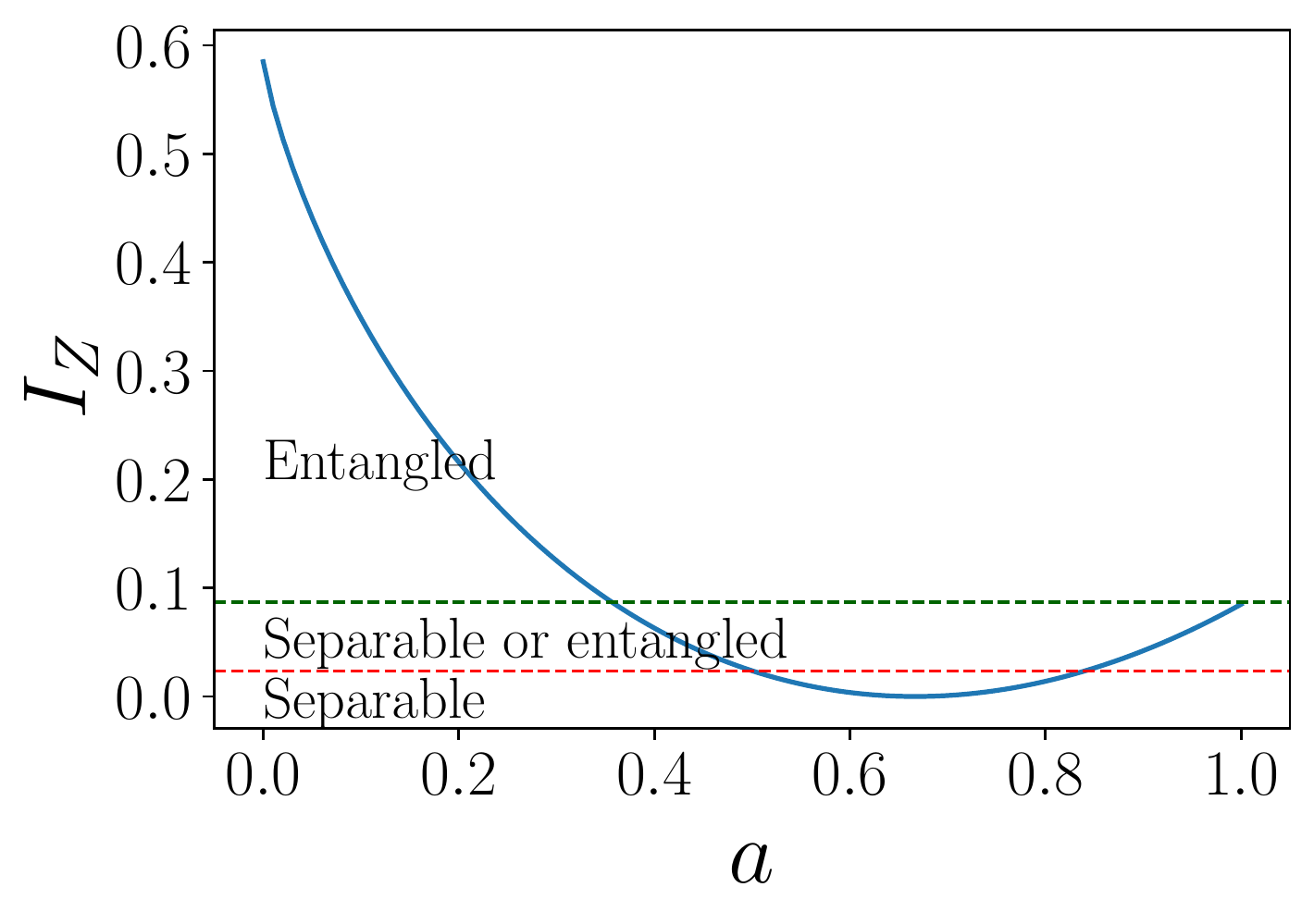}
    \includegraphics[width=1\hsize]{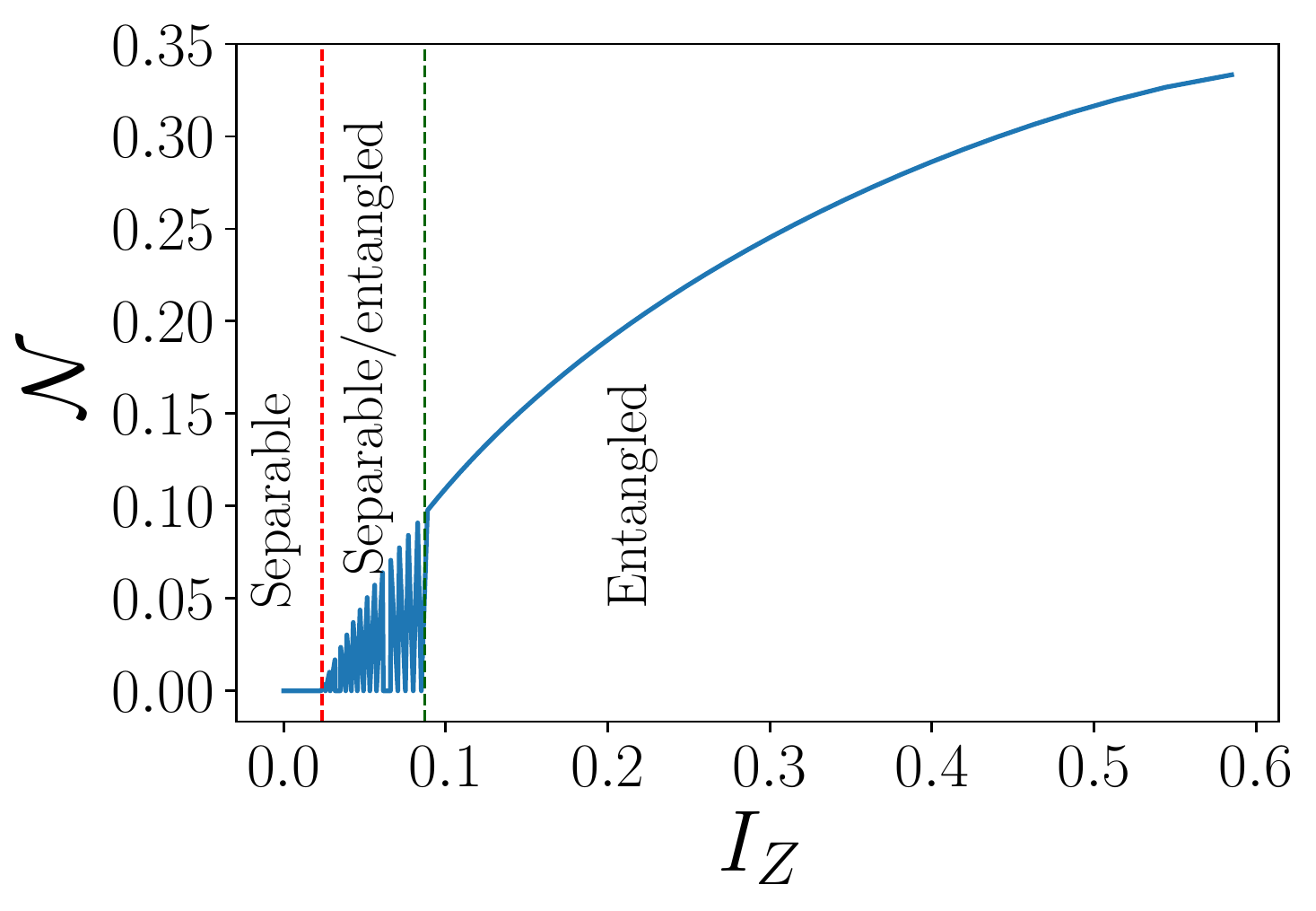}
    \caption{(Top) The plot of MI ($I_Z$) with respect to the $Z$ basis versus the parameter $a$ shows that there is a range of $I_Z$ for which the Werner state could be either entangled or separable. However, beyond a threshold, the value of $I_Z$ necessarily implies entanglement. (Bottom) The plot of Negativity ($\mathcal{N}$) of the Werner state versus $I_Z$. The plot shows that within the range of $I_Z$ for which entanglement is not guaranteed, $\mathcal{N}$ shows an oscillatory behaviour between zero and non-zero values. Beyond the threshold, $\mathcal{N}$ is a monotonic function of $I_Z$.}
    \label{fig:mi_werner}
\end{figure}

\begin{figure}[h]
\centering
\includegraphics[width=1\hsize]{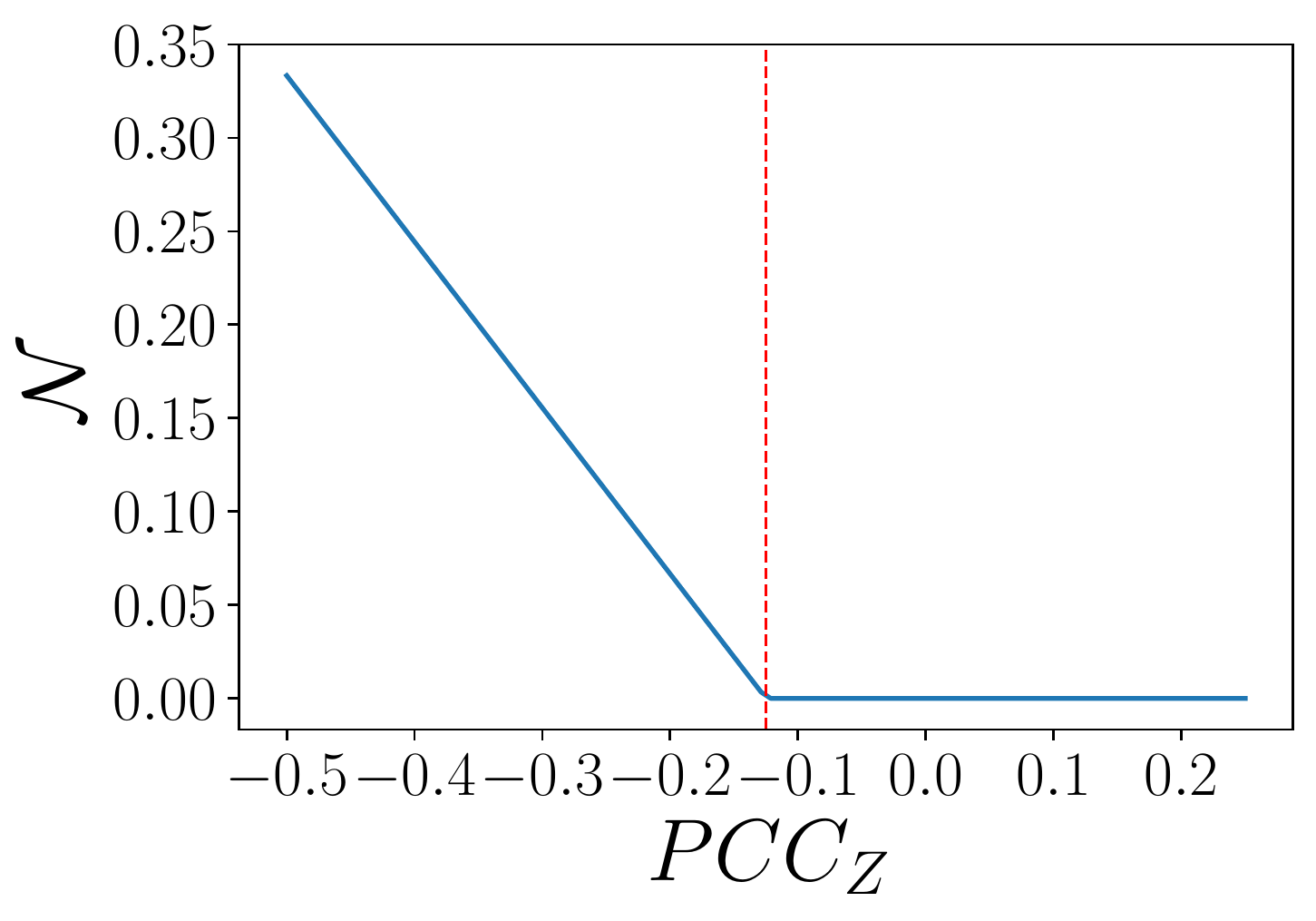}
\caption{Negativity $(\mathcal{N})$ of the Werner state as a function of PCC with respect to the $Z$-basis ($PCC_{Z}$). The plot shows that for the entangled Werner states, $PCC_Z$ has an upper-bound. Below that bound, $\mathcal{N}$ depends linearly on the value of $PCC_{Z}$.}
\label{fig:pcc_werner_comp}
\end{figure}

The density matrix of Werner state is given by
\begin{align}
    \label{eq:werner_main}
    \rho_\mathrm{w} =& a \frac{2}{d(d+1)} P_\mathrm{sym}+ (1-a) \frac{2}{d(d-1)} P_\mathrm{as}
\end{align}
where
\begin{align}
    \label{eq:psym}
    P_\mathrm{sym} =& \frac{1}{2} \left(\mathbb{1} + P \right)\\
    \label{eq:pas}
    P_\mathrm{as} =& \frac{1}{2} \left(\mathbb{1} - P \right)\\
    \label{eq:p_werner}
    P =& \sum\limits_{i,j} \ket{i}\bra{j} \otimes \ket{j}\bra{i}
\end{align}
Werner state as defined in equation \ref{eq:werner_main} is known to be entangled for $a < 1/2$ with Negativity
\begin{align}
    \label{eq:neg_werner}
    \mathcal{N}_\mathrm{w} =& \max\left\{0, \frac{1-2a}{d}\right\}
\end{align}
%
Here the joint-probability distribution of measurement outcomes with respect to the $Z$ basis is given by (see Appendix \ref{app:noisy-bell-probs} for calculations)
\begin{align}
    p_Z(i,j)=\begin{cases}
    \frac{1 - d\mathcal{N}}{d(d+1)} & i=j \\
    \frac{1 + \mathcal{N}}{d^2 - 1} & i \neq j
    \end{cases}
\end{align}
Similar to the case of the noisy Bell-state, here the statistical correlators with respect to the $Z$ basis can be written as functions of the Negativity of Werner states. In Figures \ref{fig:mp_werner_comp}, \ref{fig:mi_werner}, \ref{fig:pcc_werner_comp} we show the plots of Negativity versus the three correlators for the Werner state for $d=3$. Note that, both MP and PCC are monotonically related to the Negativity. However, there is a range of values of MI for which there is an  ambiguity in determining the Negativity of Werner states. This is because, in this range, there is no one-to-one correspondence between MI and the parameter $a$ in Eq.~\eqref{eq:werner} defining the Werner state. Beyond this region, MI can be used to determine the Negativity of entangled Werner states.

\subsection{One-parameter Horodecki state}\label{sec:horodecki}
The one-parameter Horodecki (OPH) state is another important state which showed, for the first time, the existence of non-distillable entangled states called bound entanglement \cite{Horodecki1999}. The simplest example of a bound entangled state is provided by the OPH state which is a two-qutrit state. We show that the statistical correlators can also be used to detect entanglement of such non-distillable states. Moreover, for the distillable entangled OPH states there exists a monotonic relation between the Negativity and the statistical correlators. 

The density matrix of this state is
\begin{align}
    \label{eq:horodecki_def}
    \rho_\mathrm{H} =& \frac{2}{7} \ket{\phi^+}\bra{\phi^+} + \frac{a}{7}\sigma_+ + \frac{5-a}{7} \sigma_-
\end{align}
where $\phi^+$ is the $3 \times 3$ maximally entangled state defined in equation \eqref{eq:phi_plus} and 
\begin{align}
    \label{eq:sigma_plus}
    \sigma_+ =& \frac{1}{3} \left(\ket{0,1}\bra{0,1} + \ket{1,2}\bra{1,2} +\ket{2,0}\bra{2,0}\right) \\
    \label{eq:sigma_minus}
    \sigma_- =& \frac{1}{3} \left(\ket{1,0}\bra{1,0} + \ket{2,1}\bra{2,1} +\ket{0,2}\bra{0,2}\right)
\end{align}
are mixed separable states \cite{Horodecki1998BoundActivated}. The one-parameter Horodecki state is known to have the following characteristics for $2 \leq a \leq 5$
\begin{align}
    \label{eq:horodecki_parameter}
    \rho_\mathrm{H} \text{~is~} \begin{cases} \text{separable for } 2 \leq a \leq 3 \\ \text{bound entangled for } 3 < a \leq 4 \\ \text{NPT entangled for } 4 < a \leq 5 \end{cases}
\end{align}
We observe that statistical correlations of joint local measurements can be used to detect bound entanglement in the one-parameter Horodecki states. This is achieved from the monotonicity of the statistical correlators MP, MI and PCC with respect to the state parameter $a$ whose value distinguishes between the separable, non-distillable and distillable entangled OPH states.
In the distillable-entangled region, we also relate the statistical correlators to the Negativity of the state.
Negativity of the one-parameter Horodecki state as defined in Eq.~\eqref{eq:horodecki_def} is
\begin{align}
\label{eq:neg_horodecki}
\mathcal{N}_\text{H} =& \begin{cases}
\frac{1}{28} \left(2\sqrt{41-20a+4a^2} - 10\right) & 4 < a \leq 5\\
0 & 2 \leq a \leq 4
\end{cases}
\end{align}
where, in the range $4 < a \leq 5$ the state is NPT entangled. In the NPT entangled region, the Negativity expression can be inverted to get $a$ as a function of $\mathcal{N}_\text{H}$
\begin{align}
\label{eq:a_vs_n}
    a =& \frac{1}{2} \left(5 + \sqrt{9 + 140 \mathcal{N}_\text{H} + 196 \mathcal{N}_\text{H}^2}\right), & \mathcal{N}_\text{H} > 0
\end{align}
which can be used to relate MP and MI to $\mathcal{N}$. Here, we use a relabelling scheme for the measurement basis which translates to 
\begin{align}
p(i,j) = \braket{i, j\oplus_3 k|\rho | i, j\oplus_3 k}, && i,j,k \in \{0,1,2\} 
\end{align}
The reason for this is that for $k=0$, the only contribution to MP comes from the $\ket{\phi^+}$ state whose coefficient $2/7$ is independent of $a$, and therefore, it cannot be used to determine $\mathcal{N}$. The use of such a relabelled basis with $k=1$ to calculate MP and detect bound-entanglement has been shown by \cite{hiesmayr2013complementarity}. Here we stress the point that any $k\neq 0 (\text{mod } d)$ can be used for this purposes. Table \ref{tab:mp_horodecki} shows that MP of the OPH state as a function of the parameter $a$, for different values of $k$, which confirms that the MP is a function of $a$ for any $k\neq 0 (\text{mod }d)$.

Moreover, importantly, we show, using the same relabelling scheme, that MI and PCC can also be to detect bound-entanglement as discussed below. Note that, the condition $k \neq 0$ is not necessary for using MI to detect entanglement and determine $\mathcal{N}$ in the distillable region. This is because MI is a sum of weighted logarithms over all indices and, therefore, there are non-zero contributions from all the terms in the OPH state, as can be seen from Eq.~\eqref{eq:horodecki_def}. Nevertheless, for uniformity, we calculate MI with $k \neq 0 (\text{mod }d)$ like in the cases of MP and PCC.

Here we show the example with $k=1$, and the results for other choices of $k$ are similar. Specifically, for $k=1$ when $\mathcal{N}>0$, i.e., in the NPT-region
%
\begin{align}
    p_Z(i,j) = \begin{cases}
    \frac{1}{42}\left(5 + \sqrt{9 + 140 \mathcal{N}_\text{H} + 196 \mathcal{N}_\text{H}^2}\right) & i=j \\
    \frac{2}{21} & i = j \oplus_3 1\\
    \frac{1}{42}\left(5 - \sqrt{9 + 140 \mathcal{N}_\text{H} + 196 \mathcal{N}_\text{H}^2}\right) & j = i \oplus_3 1\\
    0 & \text{otherwise}
    \end{cases}
\end{align}
is the probability expressed in terms of the Negativity of the state. Consequently, the statistical correlators can be related to $\mathcal{N}$. 

\begin{figure}[h]
    \centering
    \includegraphics[width=1\hsize]{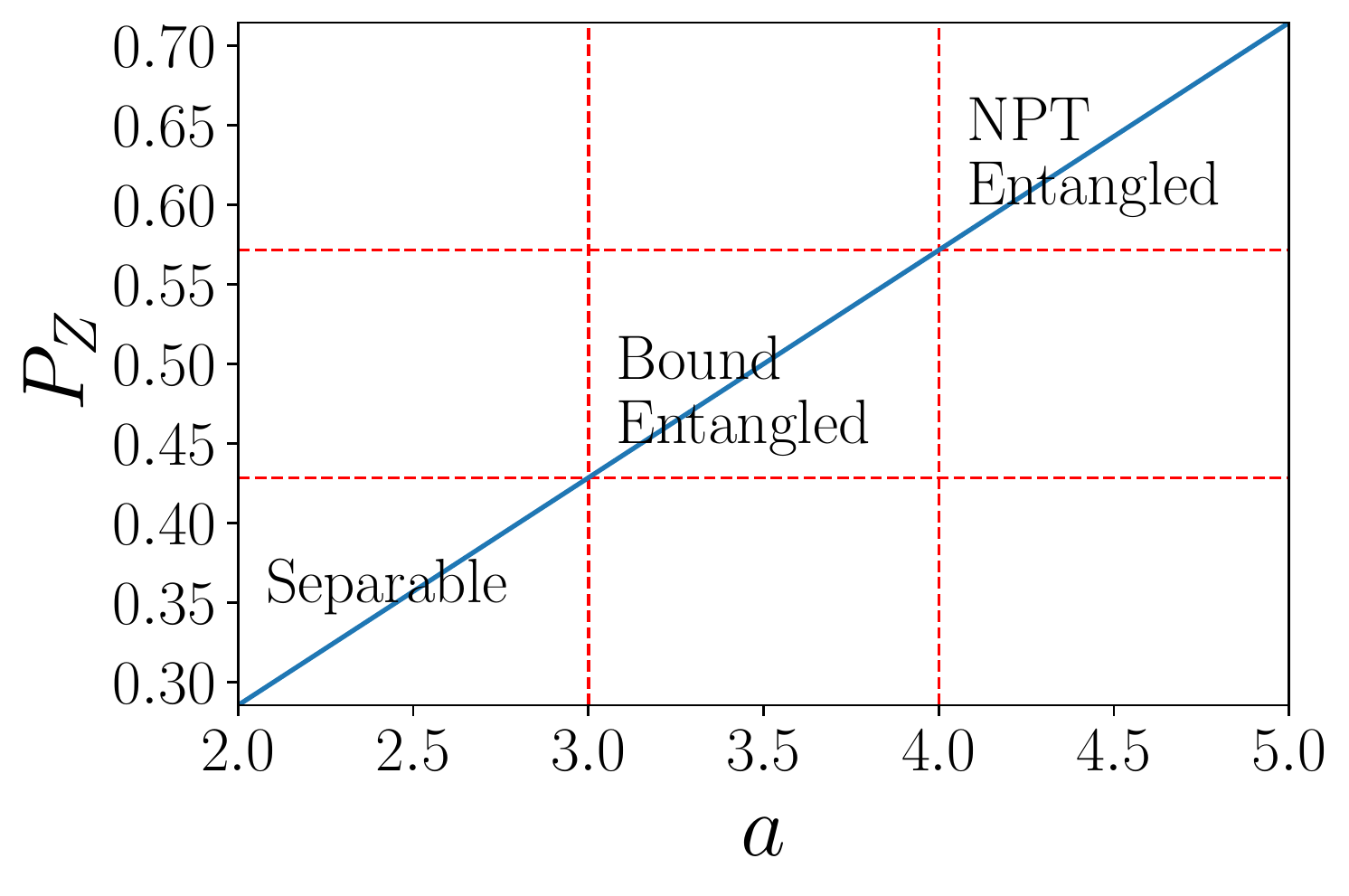}
    \caption{MP ($P_{Z}$) with respect to the $Z$ basis varies according to the values of the parameter $a$, spanning the three regions corresponding to the separable, bound entangled and NPT entangled states respectively. Therefore, it is possible to detect bound entangled OPH state from the value of $P_Z$.}
    \label{fig:mp_vs_a_horodecki}
\end{figure}

\begin{figure}[h]
    \centering
    \includegraphics[width=1\hsize]{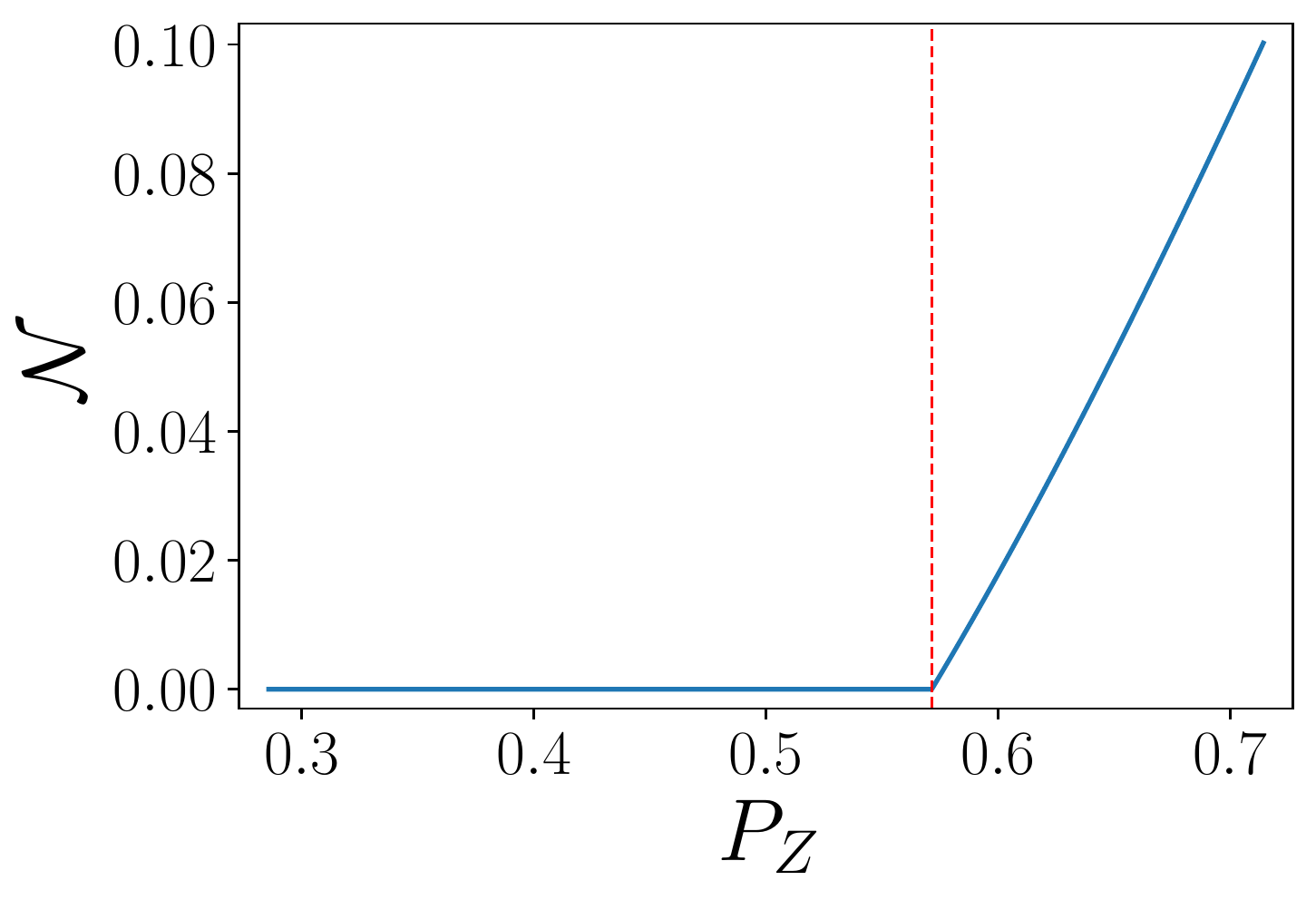}
    \caption{The Negativity $(\mathcal{N})$ of one-parameter Horodecki state as a function of MP ($P_{Z}$) with respect to the computational basis of the type $\ket{i,i\oplus_d 1}~\forall~ 0\leq i < d$. As is evident from the plot, MP acts as an entanglement witness. Moreover, in the NPT region, $\mathcal{N}$ is a monotonic (linear in this case) function of MP.}
    \label{fig:mp_ii+1_neg}
\end{figure}

\begin{figure}[h]
    \centering
    \includegraphics[width=1\hsize]{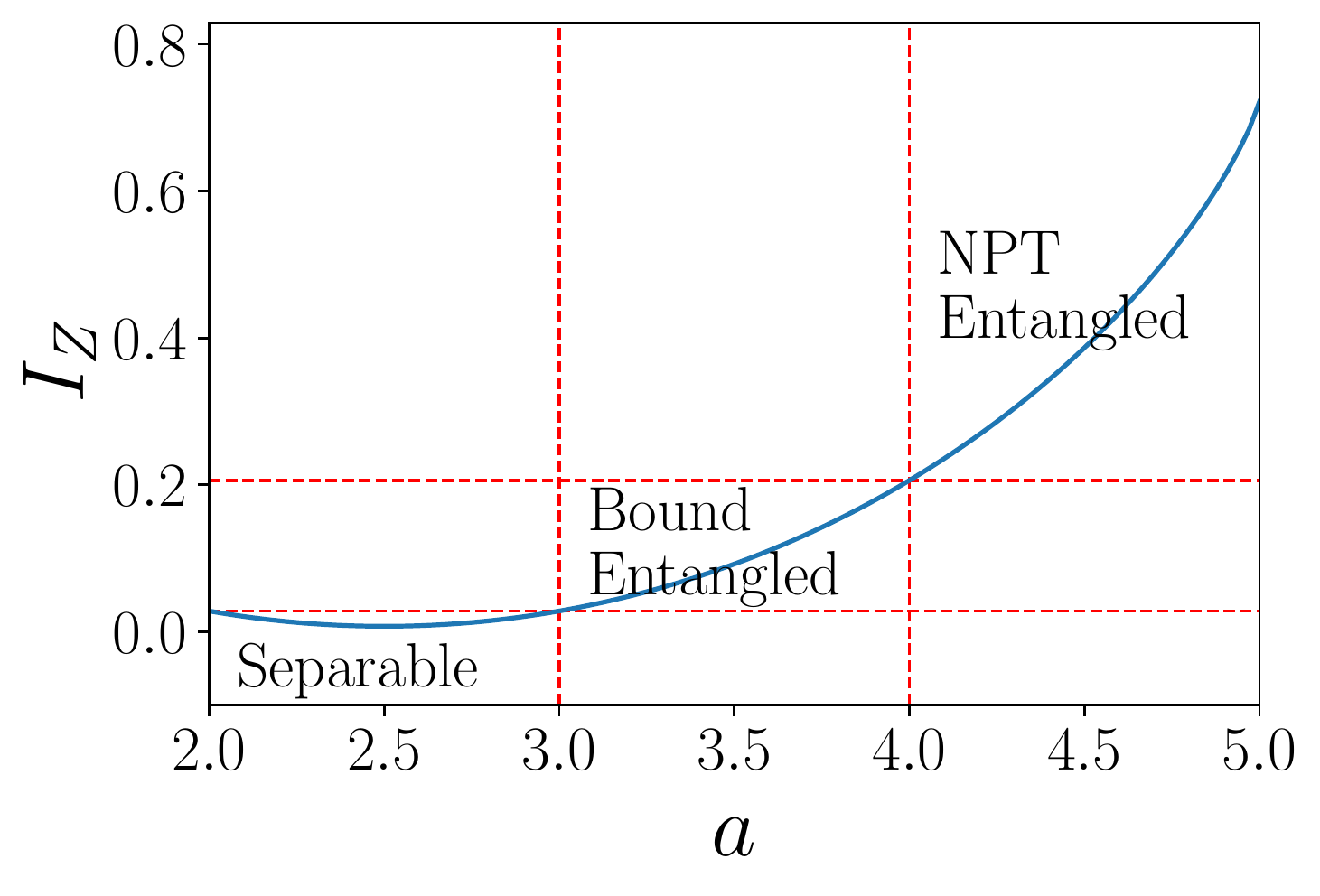}
    \caption{MI with respect to the $Z$ basis ($I_{Z}$) varies according to the values of the parameter $a$, spanning the three regions corresponding to the separable, bound entangled and NPT entangled states respectively. Therefore, it is possible to detect bound entangled OPH state from the value of $I_Z$.}
    \label{fig:mi_horodecki}
\end{figure}

\begin{figure}[h]
    \includegraphics[width=1\hsize]{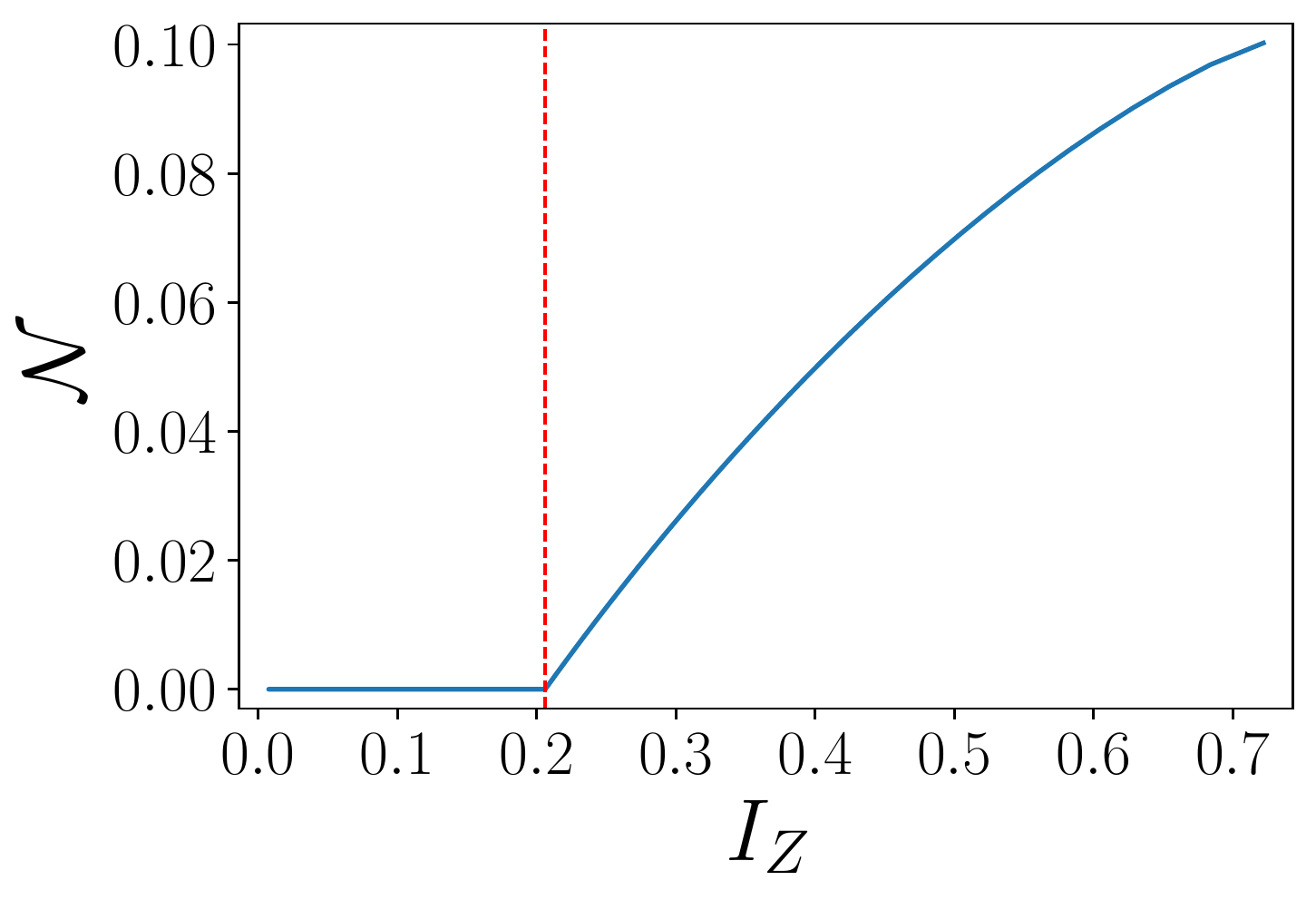}
    \caption{The Negativity $(\mathcal{N})$ of one-parameter Horodecki state as a function of MI ($I_{Z}$) with respect to the computational basis of the type $\ket{i,i\oplus_d 1}~\forall~ 0\leq i < d$. As is evident from the plot, MI acts as an entanglement witness. Moreover, in the NPT region, $\mathcal{N}$ is a monotonic (linear in this case) function of MI.}
    \label{fig:n_vs_mi__horodecki}
\end{figure}

\begin{figure}[h]
    \includegraphics[width=1\hsize]{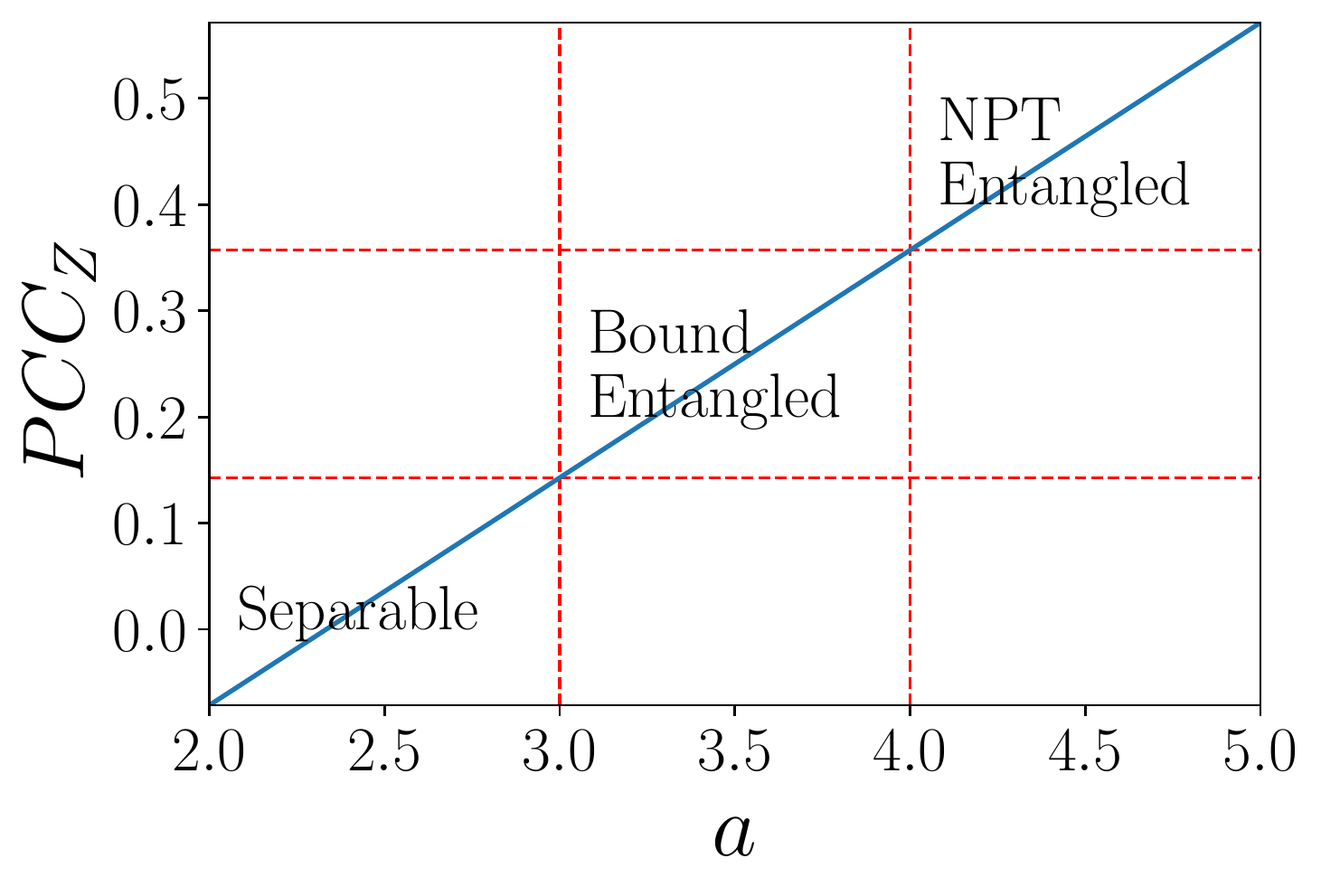}
    \caption{PCC with respect to the $Z$ basis ($PCC_Z$) varies according to the values of the parameter $a$, spanning the three regions corresponding to the separable, bound entangled and NPT entangled states respectively. Therefore, it is possible to detect bound entangled OPH state from the value of $PCC_Z$.}
    \label{fig:horo_pcc1}
\end{figure}

\begin{figure}[h]
    \includegraphics[width=1\hsize]{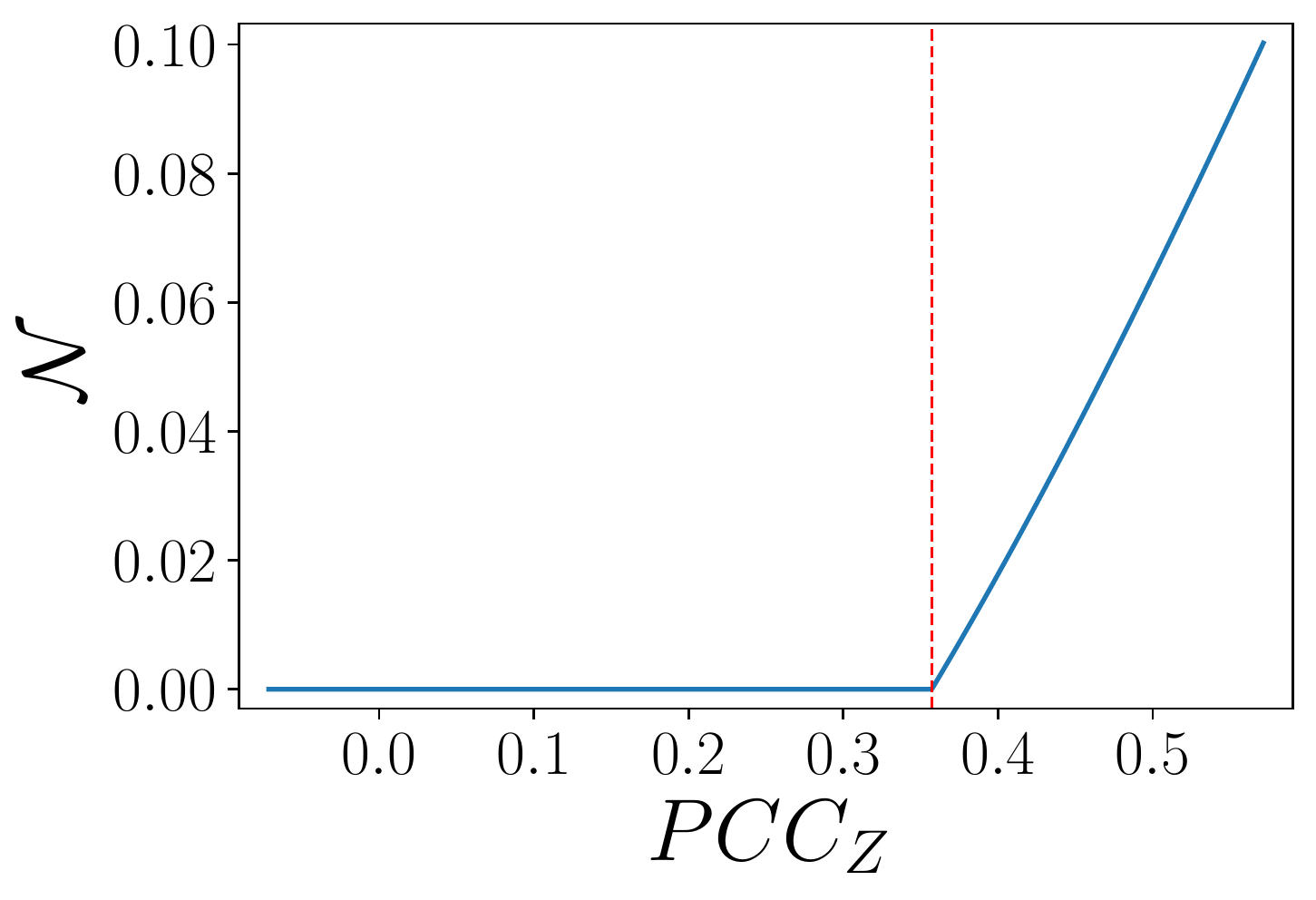}
    \caption{The Negativity ($\mathcal{N}$) of the one-parameter Horodecki state as a function of PCC with respect to the computational basis of the type $\ket{i,i\oplus_d 1}~\forall~ 0\leq i < d$. As is evident from the plot, PCC acts as an entanglement witness. Moreover, in the NPT region, $\mathcal{N}$ is a monotonic (linear) function of PCC.}
    \label{fig:horo_pcc2}
\end{figure}

We now show in the plots discussed as follows how the statistical correlators can be used to detect separable, bound-entangled and NPT-entangled states, along with getting the Negativity of the state in the distillable-region. Figure \ref{fig:mp_vs_a_horodecki} shows a plot of MP with respect to the $Z$ basis versus the parameter $a$ for the OPH states. The monotonic relation between the two quantitities makes it feasible to distinguish the separable, bound-entangled and the NPT-entangled OPH states. Note that measurements pertaining to only one basis, for example the $Z$ basis is needed for this purpose. Therefore, MP is not only useful to detect NPT-entangled states but bound-entangled states as well. In the NPT-entangled region, MP can be used to also directly measure $\mathcal{N}$, as shown in Fig.~\ref{fig:mp_ii+1_neg}. Figs. \ref{fig:mi_horodecki}, \ref{fig:n_vs_mi__horodecki}, \ref{fig:horo_pcc1} and \ref{fig:horo_pcc2} show similar results using MI and PCC for the OPH states.

Note that, owing to the monotonic relation between the statistical correlators and the state parameter of the OPH state (which determines whether the state is separable, bound-entangled or NPT entangled), one can define an ordered relation for the bound-entangled states based on the value of the statistical correlator (just like Negativity can be used to order NPT entangled states). 
{\renewcommand{\arraystretch}{1.5}
\begin{table}[]
    \centering
    \begin{tabular}{>{\centering\arraybackslash}m{0.2\hsize} >{\centering\arraybackslash}m{0.5\hsize}}
         $k$ & MP \\
         \hline
         $0$ & $2/7$\\
         $-2$ or $1$ & $a/7$\\
         $-1$ or $2$ & $(5-a)/7$
    \end{tabular}
    \caption{Mutual Predictability (MP) with respect to the $\{\ket{i, i\oplus_d k}\}$ basis as a function of the state parameter $a$ of the OPH state.}
    \label{tab:mp_horodecki}
\end{table}
}



\section{Separability bounds}\label{sec:separability_bounds}

A crucial feature of the entanglement characterization scheme formulated in this paper is that while the observables used to calculate MP, MI and PCC are fixed i.e. are independent of the state in question, the corresponding separability bounds are state-dependent functions. In particular, given the knowledge of the noises available to the experimenter, the sum of the statistical correlators in the complementary $X$ and $Z$ bases as a function of Negativity provides an upper bound for the separable states. This upper bound obtained specifically using the $X$ and $Z$ bases, is, in general, dependent on the state. In contrast, the previous studies regarding the separability bounds involving MP \cite{Spengler2012}, MI and PCC \cite{Maccone2015} yield state-independent separability bounds whose violations crucially depend on the choice of the complementary observables. Thus, there can be complementary observables for which some entangled states do not violate such bounds. These features are illustrated for the noisy-Bell and  Werner states in Appendix \ref{statedenpendentseparabilitybounds}.

\subsection{Mutual predictability based bounds}

The separability bound based on mutual predictability, formulated by Spengler et al.~\cite{Spengler2012} states that, for separable states and for any set of $m$ MUBs, the following relation holds
\begin{align}
    \sum\limits_{i=0}^{m-1} P_i \leq 1 + \frac{m-1}{d}
\end{align}
where $P_i$ is the MP for the $i^\text{th}$ MUB. This bound is necessary and sufficient for separability if the measurements are made in all $d+1$ MUBs, i.e., for $m=d+1$. For $m < (d+1)$, the separability criterion is only sufficient, but not a necessary condition for detecting entangled state. Nevertheless, with \emph{a priori} knowledge of the type of entangled state, one can find $m$ MUBs, for which the separability bound is violated. 

On the other hand, we provide an alternative approach based on the relations between the Negativity and the statistical correlators, derived in the preceding sections. We show that for a fixed choice of MUBs, the separability bound becomes state-dependent. Also, because of the monotonicity of the above mentioned relations, the separability bounds that have been obtained are necessary and sufficient to detect entanglement for the quantum states considered. The comparison between the state-independent bound of \cite{Spengler2012} and the state-dependent bounds obtained in the present study for the noisy Bell and Werner states has been discussed in Appendix \ref{statedenpendentseparabilitybounds}.

\subsection{Mutual Information based bounds}

The separability bound based on MI, formulated by Maccone et al.~\cite{Maccone2015} states that, for the separable states and for any pair of complementary observables $A\otimes B$ and $C \otimes D$, the following relation holds
\begin{align}
\label{eq:macc_mi}
    I_{AB} + I_{CD} \leq \log_2 d.
\end{align}
where $I_{AB}$ ($I_{CD}$) is MI for the observable $A\otimes B$ ($C\otimes D$). Maccone et al.~ \cite{Maccone2015} have showed that for entangled bipartite states, there exists a pair of complementary observables $A\otimes B$ and $C\otimes D$ such that the sum of MIs has a lower-bound
\begin{align}
\label{eq:macc_mi}
    I_{AB} + I_{CD} > \log_2 d.
\end{align}

Similar to the case of MP, for a given entangled state, the suitable choice of complementary observables satisfying the bound given by Eq.(\ref{eq:macc_mi}) is dependent on the state. In contrast, our approach utilizes the earlier derived monotonic relations between Negativity and MI for the fixed choice of complementary observables to obtain the necessary and sufficient separability bounds which are state-dependent. A graphical comparison between the two approaches is shown in the figures \ref{fig:compare-bound-mi-noisy} and \ref{fig:compare-bound-mi-werner} of Appendix \ref{statedenpendentseparabilitybounds}.

\subsection{Pearson Correlation Coefficient based bounds}

The separability bound based on PCC, suggested by Maccone et al.~\cite{Maccone2015} states that, for the separable states and for any pair of complementary observables $\lbrace A,C\rbrace$ and $\lbrace B,D\rbrace$, the following relation holds
\begin{align}
\label{eq:macc_pcc}
    \left|PCC_{AB}\right| + \left|PCC_{CD}\right| \leq 1.
\end{align}
where $PCC_{AB}$ ($PCC_{CD}$) denotes PCC for the observable $A\otimes B$ ($C\otimes D$). In other words, for any entangled bipartite state, there exists a pair of complementary observables $A\otimes B$ and $C\otimes D$ such that the sum of PCCs exceeds the state-independent separability bound given by Eq.(\ref{eq:macc_pcc}).\\

Similar to the results for MP and MI, for a given entangled state, the suitable choice of observables violating the separability bound  given by Eq.(\ref{eq:macc_pcc}) is dependent on the state. On the other hand, in our scheme, the monotonic relations between Negativity and PCC for the fixed choice of complementary observables yield necessary and sufficient separability bounds which are state-dependent. A graphical comparison of the two approaches is shown in the figures \ref{fig:compare-bound-mi-noisy} and \ref{fig:compare-bound-pcc-werner} of Appendix \ref{statedenpendentseparabilitybounds}.\\



While the separability bounds for MP and MI were analytically proved for an arbitrary $d$ for any bipartite state\cite{Spengler2012,Maccone2015}, the conjectured bound for PCC was proved analytically, restricted to any pure bipartite qubit state \cite{Maccone2015}. Later, it was proved for $d=3,4,5$ for a certain class of mixed states \cite{Jebarathinam2020}. For a pure bipartite qutrit state, the validity of Eq.(\ref{eq:macc_pcc}) has recently been demonstrated experimentally in \cite{g+22}.\\

As an extension of this line of work, we prove the above mentioned conjecture for the bipartite pure and coloured noise A states for an arbitrary $d$ in Appendix \ref{pearsonscorrelation}. In particular, we prove that for these states, there exists a pair of complementary observables such that $1-|PCC_{AB}| - |PCC_{CD}|\propto \mathcal{N}$, where $\mathcal{N}$ is the Negativity.\\


Thus, to summarize, there are two ways of detecting entanglement using statistical correlators: (a) Using the state-independent bounds in \cite{Spengler2012, Maccone2015} with state-dependent complementary observables. (b) Using state-dependent bounds based on our results, with the fixed choice of complementary observables. The latter scheme can be useful in the contexts where the realization of the required choice of the observables/measurement bases for implementing the scheme (a) may be difficult in a given experimental setup.

\section{Summary and outlook}\label{sec:summary}
In a nutshell, the results obtained in this study serve to comprehensively highlight the efficacy of the standard statistical correlators viz, MP, MI and PCC for characterizing the high dimensional entanglement of both distillable entangled states and non-distillable bound entangled states.
For the distillable NPT entangled states, we have considered the general case of an empirically relevant convex combination of a Bell state with the most prevalent types of noise, viz. isotropic, colored A, and colored B noises. In this case, based on the measurements for a fixed choice of at most two mutually unbiased bases (corresponding to maximum complementary observables), we have obtained the operationally useful monotonic relations between the Negativity and the statistical correlators to detect and measure entanglement. Note that, knowledge of the total amount of noise is sufficient and we need not know the amount of any individual type of noise. This gives our method an advantage over quantum state tomography in terms of the number of measurements required, provided we have the relevant information about the state considered.

The relations between the Negativity and the statistical correlators, derived in this work, with a fixed set of measurement bases, are dependent on the form of the state. This is illustrated by considering different states like the Werner state and the one parameter Horodecki state (in the \abb{npt} region). This entails knowing the type of state in order to use the relation appropriate to the state. In contrast to the separability criteria in \cite{Spengler2012, Maccone2015}, where the bounds pertaining to the criteria are state-independent but the complementary observables (or measurement bases) satisfying these criteria are state-dependent, our work implies state dependent criterion with state independent fixed choice of complementary measurement bases. The latter criterion may be particularly useful in the cases where implementing measurements of certain observables as required for applying the former criterion could be experimentally difficult. 

It should be worth probing the conceptual ramifications of the revelation that the Negativity as an entanglement measure is monotonically connected with all the standard statistical correlators which quantify correlations in the wide-ranging areas of science. Significantly, this connection is enabled by the use of measurements of only one or two complementary observables, and is valid for a range of bipartite arbitrary dimensional distillable entangled states. Therefore, these features underscore  the need for bringing out the broader fundamental significance of this line of study. For instance, what the above-mentioned connection implies regarding the physical meaning of Negativity could be instructive to probe by taking cue from the relevant earlier analysis \cite{Eltschka2013}.

Another key aspect of this paper is our demonstration that the statistical correlators can as well be used to detect bound entangled states, as shown by considering the example of the one-parameter Horodecki state. Moreover, owing to the monotonic relation between the statistical correlators and the state parameter of the OPH state (which determines whether the state is separable, bound entangled or NPT entangled), one can define an ordered relation for the bound-entangled states based on the value of the statistical correlator (just like the Negativity can be used to order the NPT entangled states). The physical implications of such an ordering of bound entangled states is a potentially interesting line of research. 

Finally, we note that the feature that our line of study enables detection of entanglement of bound OPH states using all the three standard statistical correlators MP, MI, and PCC, and can even distinguish between the bound and the NPT regimes may provide motivation for extending this direction of study. In particular, for the entanglement characterisation of, say, the  bound entangled states which show Bell nonlocality \cite{vertesi2014disproving} and steering \cite{moroder2014steering}. Such states have possible applications for quantum information tasks, for example, for extracting secure key \cite{horodecki2005secure} as well as for reducing communication complexity \cite{epping2013bound}. Thus, the use of our scheme based on statistical correlators and complementary observables could be worth exploring for complementing the entanglement witnesses \cite{yu2017family} which have been suggested for these states. 

\subsection*{Acknowledgements}
US acknowledges partial support provided by the Ministry of Electronics and Information Technology (MeitY), Government of India under grant for Centre for Excellence in Quantum Technologies with Ref. No. 4(7)/2020 - ITEA, partial support of the QuEST-DST Project Q-97 of the Govt. of India and the QuEST-ISRO research grant. DH thanks NASI for the Senior Scientist Platinum Jubilee Fellowship and acknowledges support of the QuEST-DST Project Q-98 of the Govt. of India.

\bibliographystyle{unsrt}
\bibliography{references.bib}
\newpage
\newpage
\appendix
\section{Joint probabilities of measurement outcomes for the chosen states}\label{app:noisy-bell-probs}
Statistical correlators like MP, MI and PCC are functions of the joint probabilities of outcomes of the measurements performed on the bipartite state considered. Here we explicitly derive the relevant expressions for the classes of states that we have considered.

\subsection{Noisy Bell-state}
The density operators for the three types of noises are given in equations \eqref{eq:isotropic_noise}, \eqref{eq:colourA} and \eqref{eq:colourB}. The joint-probabilities of measurement outcomes for all the three types with respect to the $Z$ basis are 
\begin{align}
    p_{\mathrm{iso},Z} =& \braket{i,j|\frac{\mathbb{1}}{d^2}|i,j} = \frac{1}{d^2}\\
    p_{\mathrm{cna},Z} =&  \frac{1}{d}\sum\limits_{a=0}^{d-1} \braket{i,j |a,a}\braket{a,a  |i,j} = \frac{1}{d}\delta_{ij}\\
    p_{\mathrm{cnb},Z} =& \frac{1}{d(d-1)} \sum\limits_{\substack{a,b=0\\a\neq b}} \braket{i,j|a,b} \braket{a,b|i,j} = \frac{1}{d(d-1)} (1 - \delta_{ij})
\end{align}
and therefore, the mutual predictabilities of all these three states are given by
\begin{align}
    P_{\mathrm{iso},Z} =&  \frac{1}{d}\\
    P_{\mathrm{cna},Z} =&  1 \\
    P_{\mathrm{cnb},Z} =& 0
\end{align}
Similarly, with respect to the $X$ basis, one obtains
\begin{align}
    p_{\mathrm{iso},X} =& \braket{i,j|\frac{\mathbb{1}}{d^2}|i,j} = \frac{1}{d^2}\\
    p_{\mathrm{cna},X} =&  \frac{1}{d}\sum\limits_{a=0}^{d-1} \braket{i,j |a,a}\braket{a,a  |i,j} =  \frac{1}{d^2}\\
    p_{\mathrm{cnb},X} =& \frac{1}{d(d-1)} \sum\limits_{\substack{a,b=0\\a\neq b}} \braket{i,j|a,b} \braket{a,b|i,j} = \frac{1}{d^2}
\end{align}
and, therefore, the mutual predictability of all these noises separately is $1/d$.

\subsection{Werner State}
\begin{align}
    \label{eq:werner}
    \rho_\mathrm{w} = & a \frac{2}{d(d+1)} P_\mathrm{sym}+ (1-a) \frac{2}{d(d-1)} P_\mathrm{as}
\end{align}
where
\begin{align}
    \label{eq:psym}
    P_\mathrm{sym} =& \frac{1}{2} \left(\mathbb{1} + P \right)\\
    \label{eq:pas}
    P_\mathrm{as} =& \frac{1}{2} \left(\mathbb{1} - P \right)\\
    \label{eq:p_werner}
    P =& \sum\limits_{i,j} \ket{i}\bra{j} \otimes \ket{j}\bra{i}
\end{align}

The joint-probability of outcomes in the $Z$ basis for the operator $P$ is
\begin{align}
    p_{P,Z} =& \sum\limits_{a,b} \braket{i,j|a,b}\braket{b,a|i,j} = \delta_{ij}
\end{align}
Note that due to the $U \otimes U$ invariance of Werner states, the probability of measurement outcomes with respect to the $X$ basis is the same as that with respect to the $Z$ basis.

\section{State-dependent separability bounds}\label{statedenpendentseparabilitybounds}
As discussed in \S\ref{sec:separability_bounds}, with the fixed choice of measurement bases, which in this work are the $X$ and $Z$ bases, our separability bounds are state dependent. This is because the relations between the Negativity and the statisitcal correlators are state-dependent. On the other hand, with the chosen bases in this paper, the separability criteria of \cite{Spengler2012,Maccone2015} are not suitable to detect all the entangled states (among the classes of states considered here). We show this by using the sums of statistical correlators with respect to the $Z$ and $X$ bases in the figures described below. For comparison, we also indicate in these figures the respective bounds of \cite{Spengler2012,Maccone2015}.\\

\begin{figure}[H]
    \includegraphics[width=1\hsize]{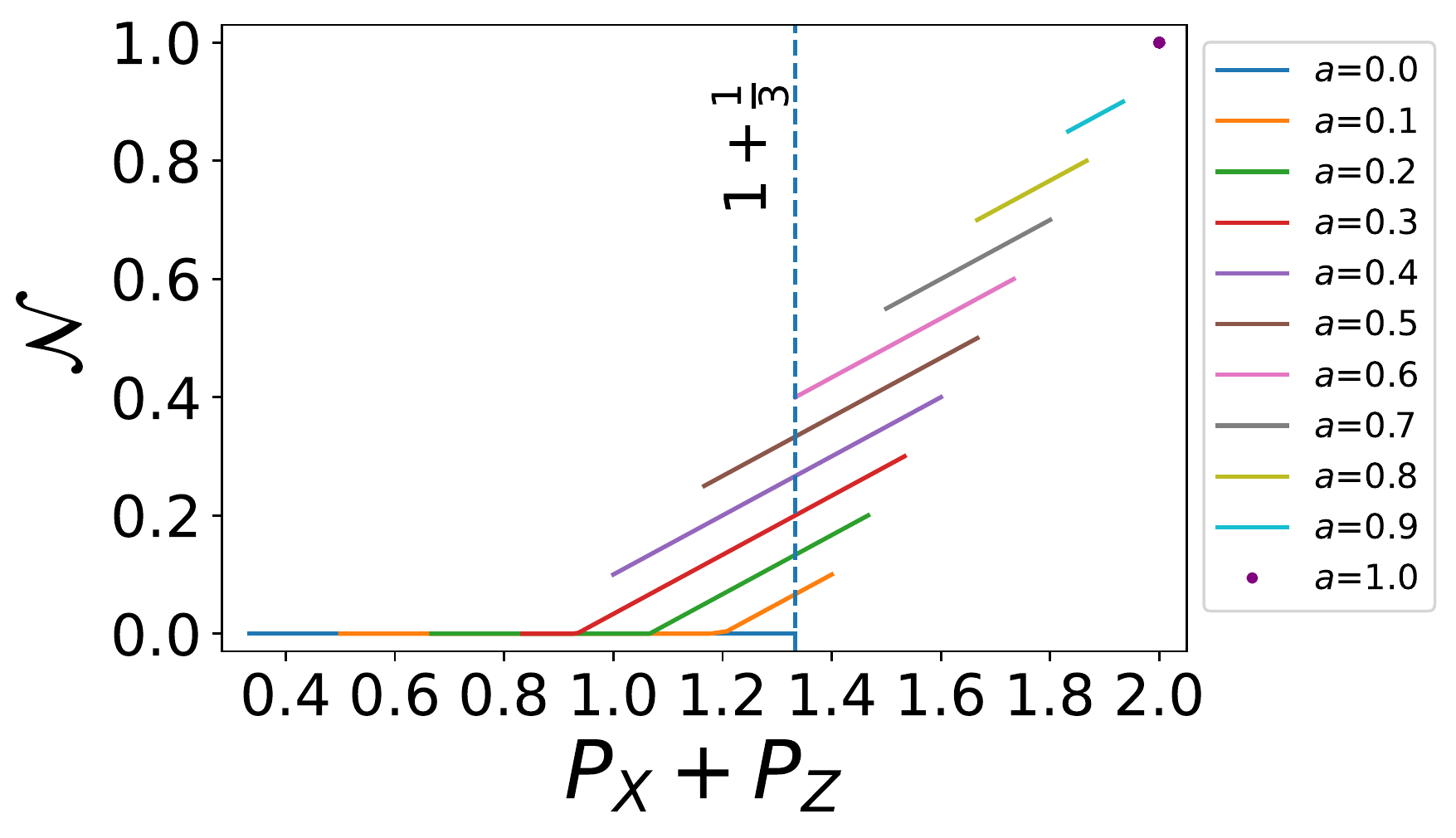}
    \caption{The Negativity $(\mathcal{N}$) of the noisy Bell state versus the sum of MPs $(P_{Z}+P_{X})$ with respect to the $X$ and $Z$ bases for $d=3$. The vertical dotted line corresponds to the state independent but observable dependent separability bound $1 + 1/d$. The line for each value of the state parameter $a$ intersects the horizontal line corresponding to $\mathcal{N}=0$, with the point of intersection indicating the state dependent separability bound for a given value of $a$.}
    \label{fig:compare-bound-mp-noisy}
\end{figure}

\begin{figure}[H]
    \includegraphics[width=1\hsize]{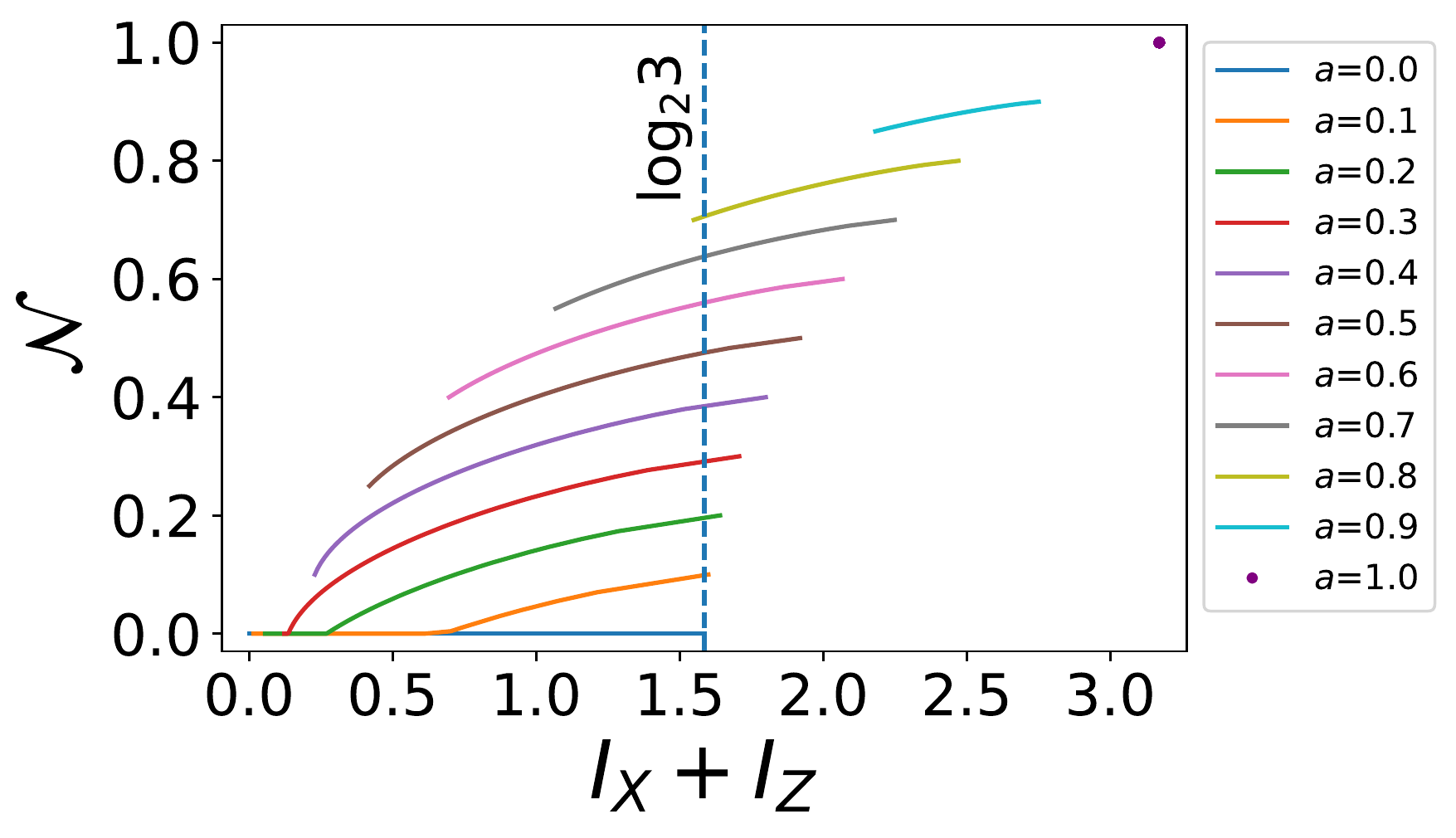}
    \caption{The Negativity $(\mathcal{N}$) of the noisy Bell state versus the sum of MIs $(I_{Z}+I_{X})$ with respect to the $X$ and $Z$ bases for $d=3$. The vertical dotted line corresponds to the state independent but observable dependent separability bound $\log_2 d$. The line for each value of the state parameter $a$ intersects the horizontal line corresponding to $\mathcal{N}=0$, with the point of intersection indicating the state dependent separability bound for a given value of $a$.}
    \label{fig:compare-bound-mi-noisy}
\end{figure}

\begin{figure}[H]
    \includegraphics[width=1\hsize]{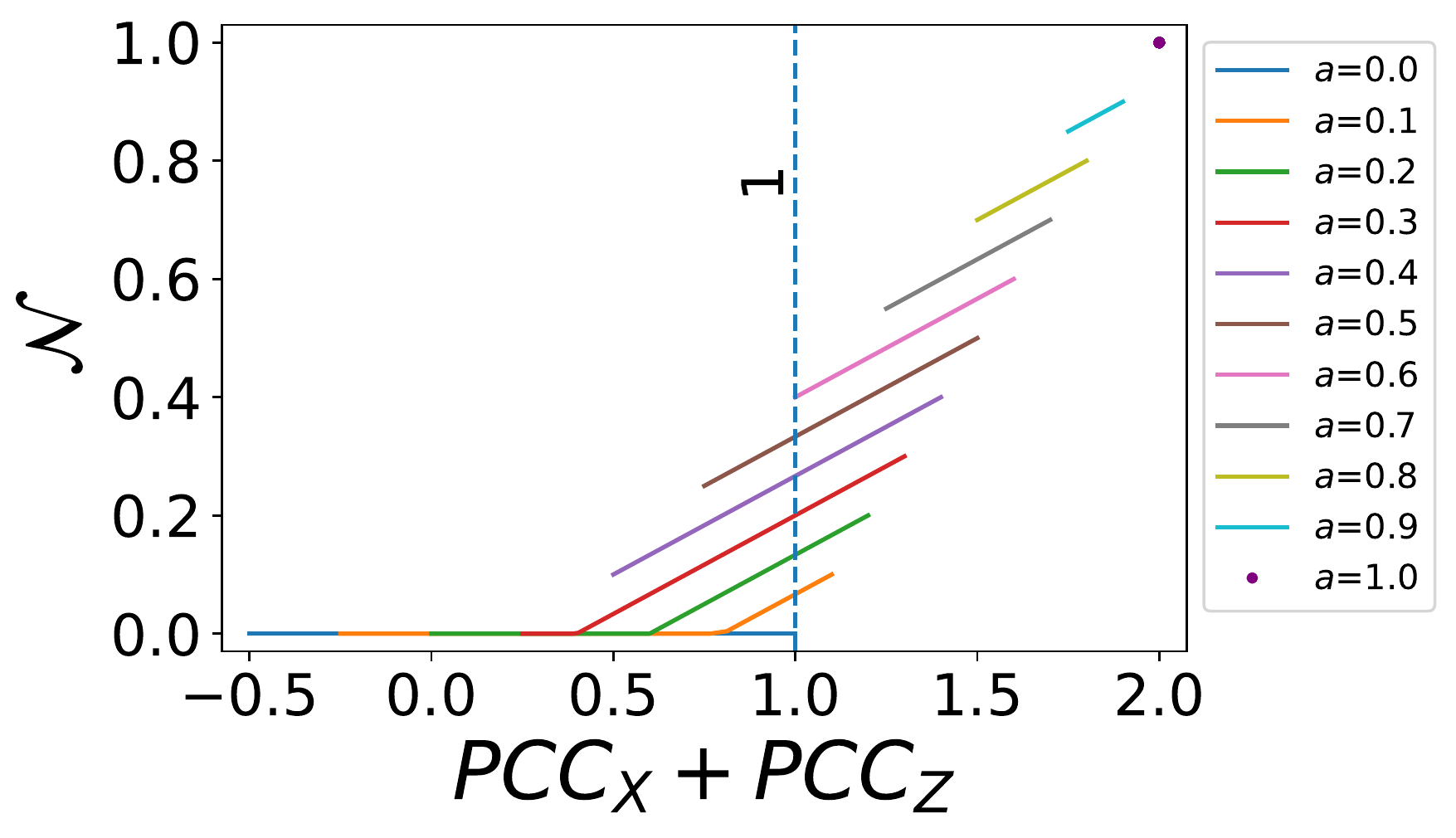}
    \caption{The Negativity $(\mathcal{N})$ of the noisy Bell state versus the sum of PCCs $(PCC_{X}+PCC_{Z})$ with respect to the $X$ and $Z$ bases for $d=3$. The vertical dotted line corresponds to the state independent but observable dependent separability bound $1$. The line for each value of the state parameter $a$ intersects the horizontal line corresponding to $\mathcal{N}=0$, with the point of intersection indicating the state dependent separability bound for a given value of $a$.}
    \label{fig:compare-bound-pcc-noisy}
\end{figure}

\begin{figure}[H]
    \includegraphics[width=1\hsize]{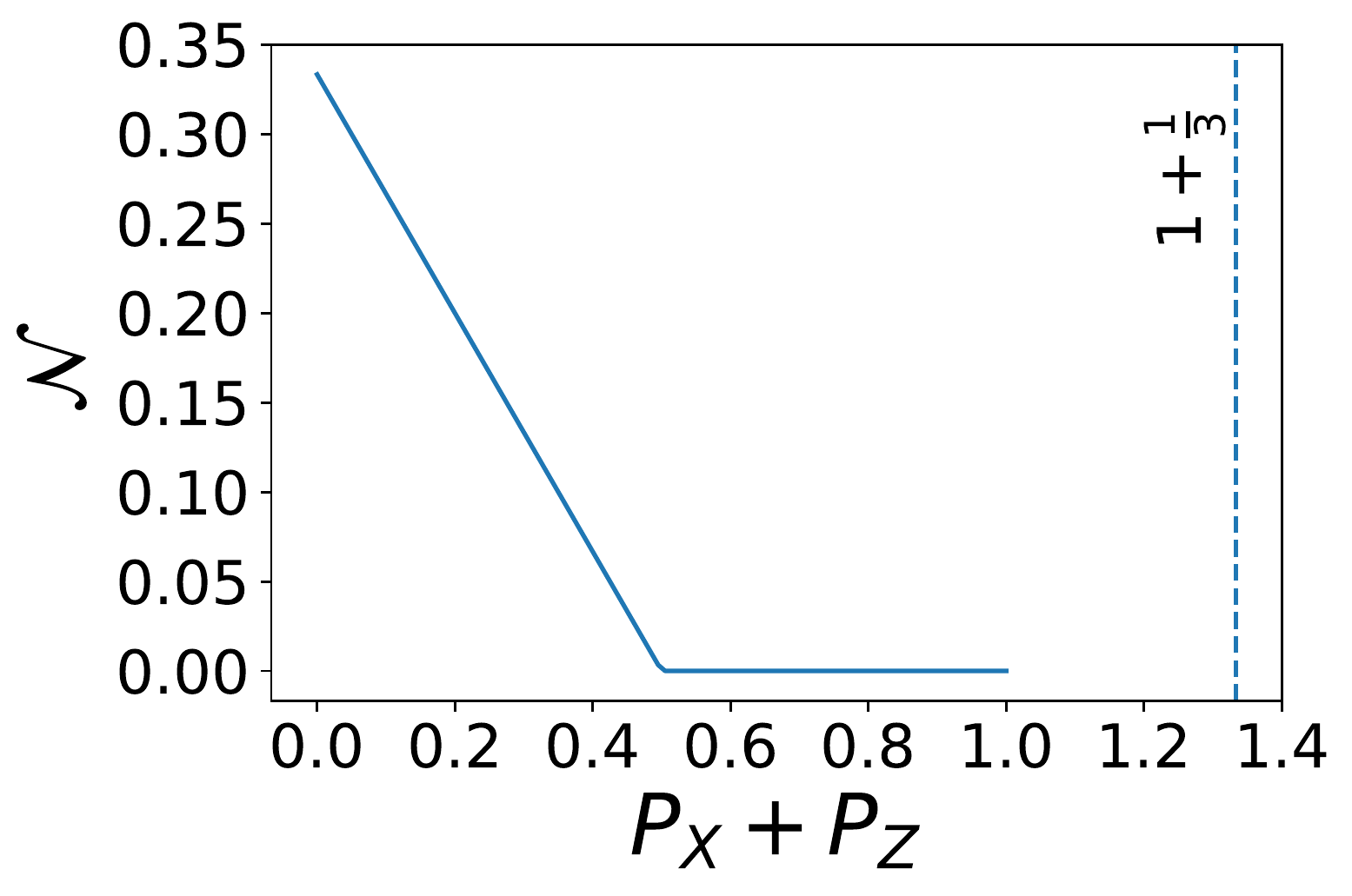}
    \caption{The Negativity $(\mathcal{N})$ of the Werner state versus the sum of MPs $(P_{Z}+P_{X})$ with respect to the $X$ and $Z$ bases for $d=3$. The vertical dotted line corresponds to the state independent but observable dependent separability bound $1 + 1/d$.}
    \label{fig:compare-bound-np-werner}
\end{figure}

\begin{figure}[H]
    \includegraphics[width=1\hsize]{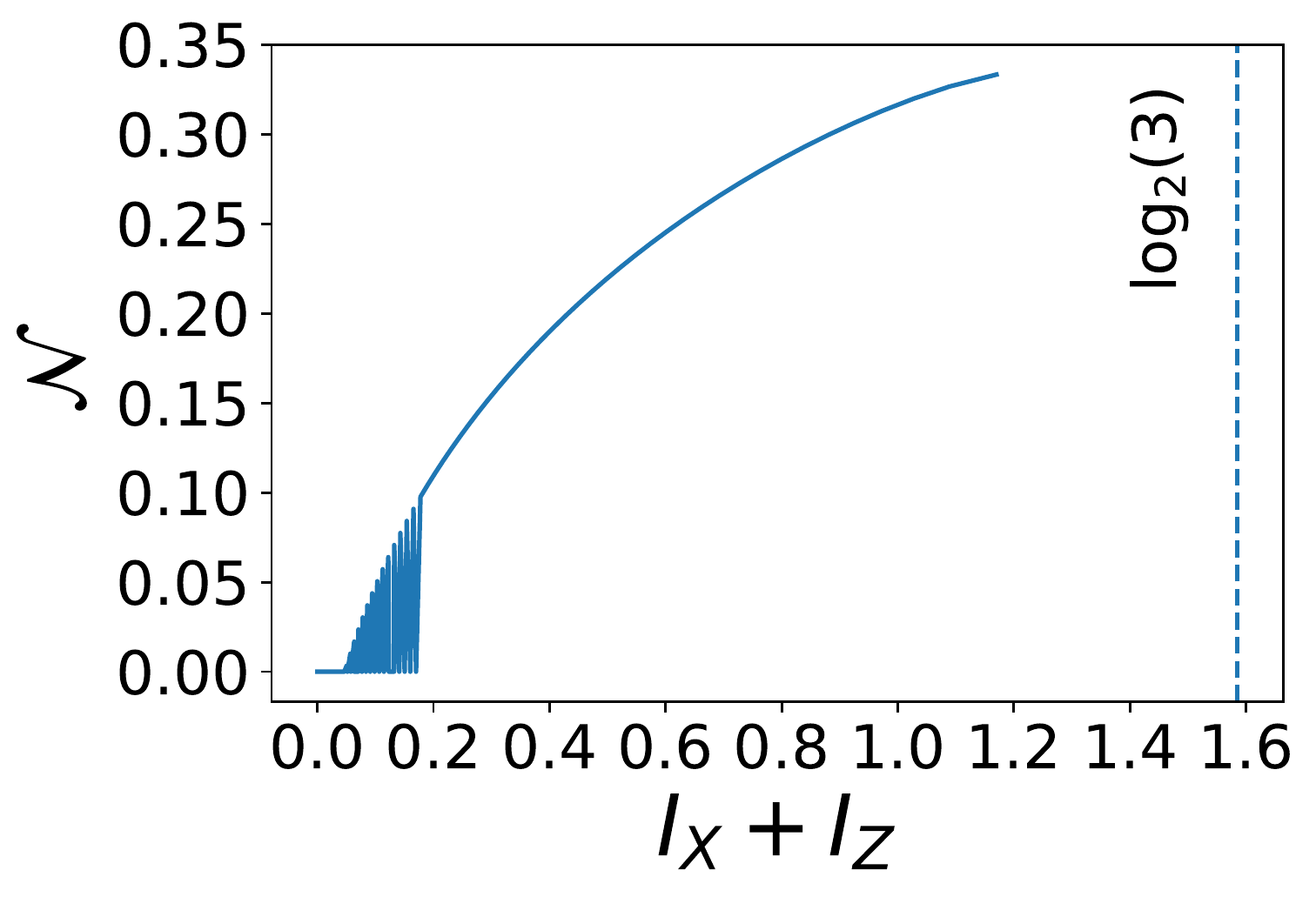}
    \caption{The Negativity $(\mathcal{N})$ of the Werner state versus the sum of MIs $(I_{Z}+I_{X})$ with respect to the $X$ and $Z$ bases for $d=3$. The vertical dotted line corresponds to the state independent but observable dependent separability bound $\log_2 d$.}
    \label{fig:compare-bound-mi-werner}
\end{figure}

\begin{figure}[H]
    \includegraphics[width=1\hsize]{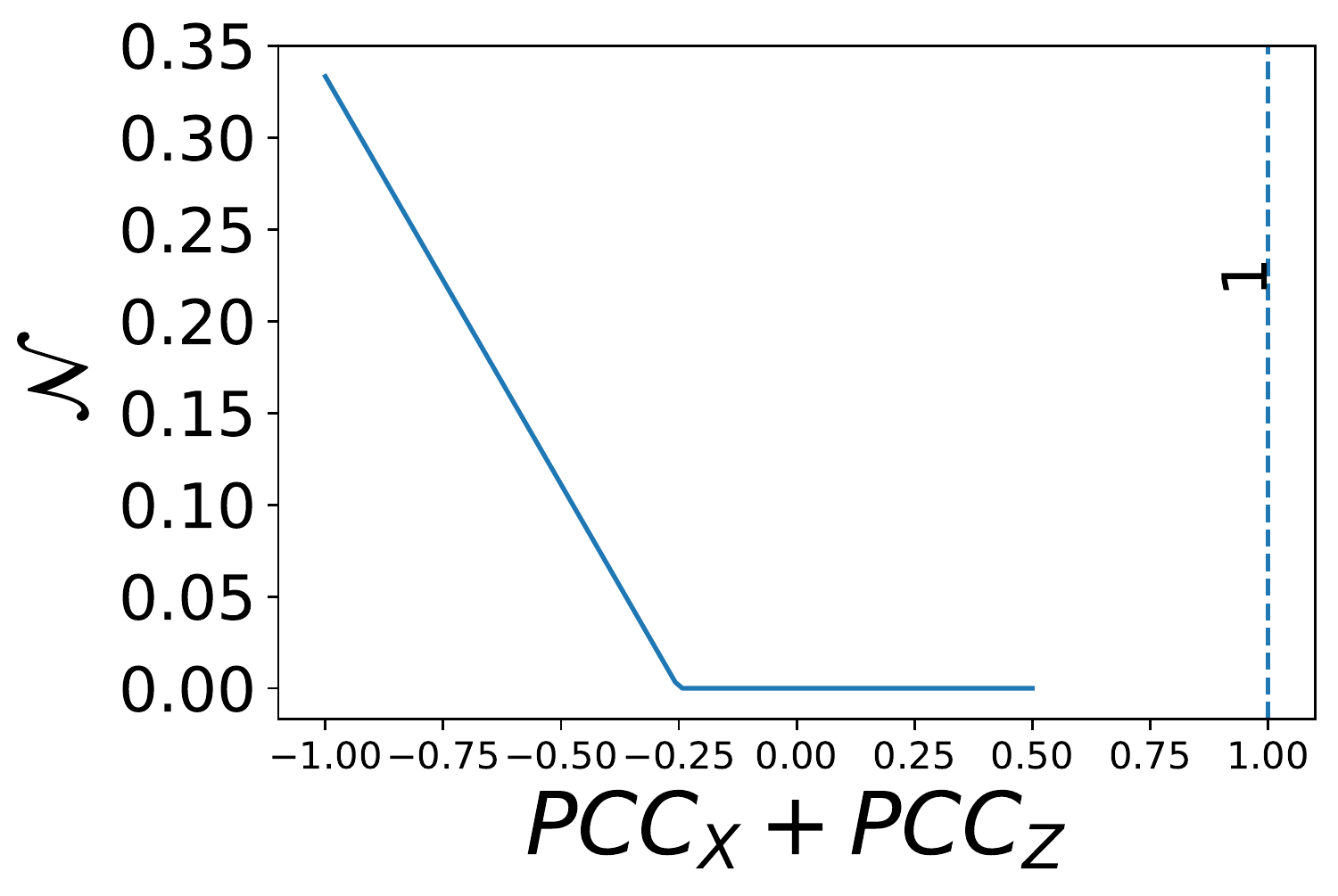}
    \caption{The Negativity $(\mathcal{N})$ of the Werner state versus the sum of PCCs $(PCC_{X}+PCC_{Z})$ with respect to the $X$ and $Z$ bases. The vertical dotted line corresponds to the state independent but observable dependent separability bound given by $1$.}
    \label{fig:compare-bound-pcc-werner}
\end{figure}

In the figures \ref{fig:compare-bound-mp-noisy},\ref{fig:compare-bound-mi-noisy},\ref{fig:compare-bound-pcc-noisy} we have plotted the Negativity of the Noisy Bell state given by eq.(\ref{eq:neg_noisy_bell}) with respect to the sums of MPs, MIs and PCCs for the $X$ and $Z$ bases respectively in the case of $d=3$. Note that, for all the three statistical correlators, the corresponding state-dependent separability bounds depend upon the mixedness parameter $a$ as indicated by eqs.(\ref{eq:joint_prob_noisy_bell}) and (\ref{eq:joint_prob_noisy_bell2}). In the figures \ref{fig:compare-bound-np-werner},\ref{fig:compare-bound-mi-werner} and \ref{fig:compare-bound-pcc-werner} we have plotted the Negativity of Werner state with respect to the sums of MPs, MIs and PCCs for the $X$ and $Z$ bases respectively in the case of $d=3$. For these observables, there exists entangled states such that the sums of the statistical correlators are less than the state-independent bounds. \cite{Spengler2012,Maccone2015} In other words, the use of the observables $X$ and $Z$ is not suitable to detect certain entangled states using the state-independent bounds but which can be detected using the state-dependent bounds.


\section{Proof of Maccone et al. Conjecture for pure and colored noise A states}\label{pearsonscorrelation}

Here we deal with the problem of finding pairs of complementary observables $\lbrace \hat{A},\hat{C} \rbrace$ and $\lbrace \hat{B},\hat{D} \rbrace$ such that the sum of \abb{pcc}s $|PCC_{AB}|+|PCC_{CD}|>1$ for a bipartite arbitrary dimensional entangled state. For two given observables, say, $X$ and $Y$, $PCC_{XY}$ is given by

\begin{equation}
\label{eq:pccexpr}
PCC_{XY} = \frac{\langle X\otimes Y\rangle-\langle X\otimes\mathbb{1}\rangle\langle\mathbb{1}\otimes Y\rangle}{\sqrt{\langle X^{2}\otimes\mathbb{1}\rangle-\langle X\otimes\mathbb{1}\rangle^{2}}\sqrt{\langle\mathbb{1}\otimes Y^{2}\rangle-\langle\mathbb{1}\otimes Y\rangle}}
\end{equation}

Taking the Negativity ($\mathcal{N}$) as the measure of entanglement, we show that, using suitable pairs of complementary observables $\{A,C\}$ and $\{B,D\}$, $|PCC_{AB}|+|PCC_{CD}|-1\propto \mathcal{N}$ for certain types of states, viz, pure and colored noise A state.

We proceed as follows. First, we construct the relevant observables and demonstrate their mutual unbiasedness. Next, we obtain the values of the relevant Pearson correlators for a $d$ dimensional bipartite state. Finally, we derive the desired relationship between the sum of Pearson correlators and the Negativity for pure  and colored noise A states.

\subsection{Construction of the required observables and their properties}

The complementary observables for each subsystem are the generalized $Z$ observable and a modification of the generalized $X$ observable which is given by

\begin{equation}
 \label{eq:obsexpr1}
W = \sum_{i,j:\braket{i|j}= 0}\ket{j}\bra{i}
\end{equation}

In eq.(\ref{eq:obsexpr1}) the summation is over both $i$ and $j$. For a given $i$, we sum over only those values of $j$ which are orthogonal to $i$, i.e $\braket{i|j}=0$. Another way to express the above observable is to consider generalized Pauli basis $\hat{\sigma}_{j,k}=\ket{j}\bra{k}+\ket{k}\bra{j}$ with $j>k$. We can then write $W=\underset{j,k}{\sum}\hat{\sigma}_{j,k}$. \\

The observable defined in eq.(\ref{eq:obsexpr1}) projects each computational basis vector $\ket{j}$ to $\sum_{k:\braket{k|j}=0}\ket{k}$ i.e. to its complete orthogonal subspace, $|i\rangle\longrightarrow\sum_{j:\langle i|j\rangle=0}|j\rangle$.\\

Now, we show that the observables $Z$ and $W$ are complementary to each other i.e. the corresponding eigenstates are mutually unbiased. 

\subsubsection*{Demonstration of maximum complementarity of the $W$ and $Z$ observables}

Note that the eigenstates of $Z$ form the computational basis $\{|j\rangle\}$. Therefore, to show the maximum complementarity of $Z$ and $W$ we only need to show that the eigenvectors of $W$ are mutually unbiased(MUB) to the states of the computational basis. For this purpose, we define a suitable basis, mutually unbiased with respect to the computational basis $\{|j\rangle\}$ given by $\lbrace|k\rangle=\frac{1}{\sqrt{d}}\sum_{j}\E^{\frac{2\I\pi jk}{d}}\ket{j}\rbrace$.\\
    
    For $k=0$ we can show that
\begin{align}
W\ket{k=0} & = \frac{1}{\sqrt{d}}\sum_{j}W\ket{j}\\
& =\frac{1}{\sqrt{d}}\sum_{j,k:\braket{k|j}=0}\ket{k}\\
& =\frac{(d-1)}{\sqrt{d}}\sum_{k}\ket{k}\\
&=(d-1)\ket{k=0}
\end{align}

The above calculation shows us that $\ket{k=0}$ is one of the eigenstates of $W$ with eigenvalue $(d-1)$. For $1 \leq k \leq (d-1)$, taking $\omega=\E^{\frac{2\I\pi}{d}}$ we obtain
\begin{align}
W\ket{k} & = \frac{1}{\sqrt{d}}\sum_{j}\omega^{jk}W\ket{j} \nonumber\\
& =\frac{1}{\sqrt{d}}\sum_{j,l:\braket{l|j}=0}\omega^{jk}\ket{l}\\
\label{eq:wzmub}
&=\frac{1}{\sqrt{d}}\sum_{l}\ket{l}\sum_{j:\braket{l|j}=0}\omega^{jk}
\end{align}
Note that, for any $k$ which is not a multiple of $d$ we have 
\begin{equation}
\label{eq:complexid}
    \sum_{j=0}^{d-1}\omega^{jk}=\frac{1-\E^{2\I\pi k}}{1-\E^{\frac{2\I\pi k}{d}}}=0.
\end{equation}
Using Eq.~(\ref{eq:complexid}) we can rewrite Eq.(\ref{eq:wzmub}) as
\begin{align*}
W\ket{k} &=-\frac{1}{\sqrt{d}}\sum_{l}\ket{l}\omega^{lk}\\
& =-\ket{k}
\end{align*}
In other words, the states $\ket{k}=\frac{1}{\sqrt{d}}\sum_{j}\omega^{jk}\ket{j}$ with $1\leq k\leq (d-1)$ are the eigenstates of $W$ with eigenvalues $-1$.\\

Next, we outline the steps used in proving a relation which will be crucial for calculating the denominator of the expression of PCC given by Eq.~(\ref{eq:pccexpr}). \\

\begin{itemize}
\item The following relation holds for the observable given by Eq.(\ref{eq:obsexpr1})
\begin{equation}
\label{eq:sqW}
W^{2}=\left[(d-1)\mathbb{1}+(d-2)W\right]
\end{equation}
Note that, from Eq.(\ref{eq:obsexpr1}) it is evident that we can write the matrix form of the observable $W$ as follows:
\begin{equation}
\label{hwo2}
W=\begin{bmatrix} 
    0 & 1 & 1 & 1 & 1 & \dots & 1 \\
1 & 0 & 1 & 1 & 1 & \dots & 1 \\
1 & 1 & 0 & 1 & 1 & \dots & 1 \\
1 & 1 & 1 & 0 & 1 & \dots & 1 \\
1 & 1 & 1 & 1 & 0 & \dots & 1 \\
\vdots & \ddots & \\
1 & 1 & 1 & 1 & 1 & \dots & 0 
    \end{bmatrix}
\end{equation}
Consequently, for $W^{2}$ we have the following expression
\begin{align}
W^{2} & =\begin{bmatrix}
 0 & 1 & 1 & 1 & 1 & \dots & 1 \\
1 & 0 & 1 & 1 & 1 & \dots & 1 \\
1 & 1 & 0 & 1 & 1 & \dots & 1 \\
1 & 1 & 1 & 0 & 1 & \dots & 1 \\
1 & 1 & 1 & 1 & 0 & \dots & 1 \\
\vdots & \ddots & \\
1 & 1 & 1 & 1 & 1 & \dots & 0 
\end{bmatrix}
\begin{bmatrix}
 0 & 1 & 1 & 1 & 1 & \dots & 1 \\
1 & 0 & 1 & 1 & 1 & \dots & 1 \\
1 & 1 & 0 & 1 & 1 & \dots & 1 \\
1 & 1 & 1 & 0 & 1 & \dots & 1 \\
1 & 1 & 1 & 1 & 0 & \dots & 1 \\
\vdots & \ddots & \\
1 & 1 & 1 & 1 & 1 & \dots & 0 
\end{bmatrix}\nonumber\\
& = \begin{bmatrix} 
    (d-1) & (d-2) & (d-2) & (d-2) & \dots & (d-2) \\
(d-2) & (d-1) & (d-2) & (d-2) & \dots & (d-2) \\
(d-2) & (d-2) & (d-1) & (d-2) & \dots & (d-2) \\
(d-2) & (d-2) & (d-2) & (d-1) & \dots & (d-2) \\
\vdots & \vdots & \vdots & \vdots & \ddots & \vdots\\
(d-2) & (d-2) & (d-2) & (d-2) & \dots & (d-1) \\
    \end{bmatrix}\nonumber\\
    \label{sqhwo}
& =(d-1)\mathbb{1}+(d-2)W
\end{align}

\end{itemize}

\subsection{Pearson correlators for a $d$ dimensional bipartite state}

Here we find the values of $PCC_{Z}$ and $PCC_{W}$ for a general bipartite qudit state.\\

To set the stage, we will define a few mathematical notations. The vector $|\bar{i}\rangle$ denotes an arbitrary vector from the computational basis spanning the orthogonal subspace of $\ket{i}$. For example, for $d=3$, and the computational basis $\{|0\rangle,|1\rangle,|2\rangle\}$, the vector $|\bar{0}\rangle$ would denote an arbitrary vector from the set $\{|1\rangle,|2\rangle\}$. We use this notation in Eq.~(\ref{eq:obsexpr1}). For a fixed $i$, the summation $\sum_{j:\langle i|j\rangle=0}$ is equivalent to $\sum_{\bar{i}}$. In this notation, we can re-write Eq.(\ref{eq:obsexpr1}) as follows:
\begin{equation}
\label{eq:obsexpr2}
W =\sum_{i,\bar{i}}|\bar{i}\rangle\langle i|
\end{equation}

Given a $d_{a}\times d_{b}$ bipartite state
\begin{equation}
\label{eq:genbiparqudit}
\rho_{ab}=\underset{i,j;k,l}{\sum} \epsilon_{i,j;k,l}|i\rangle\langle j|\otimes|k\rangle\langle l|,
\end{equation}
for $W_{a}=\sum_{i,\bar{i}}|\bar{i}\rangle\langle i|$ and $W_{b}=\sum_{k,\bar{k}}|\bar{k}\rangle\langle k|$, the following relationships hold
\begin{align}
\label{numPCC1}
\langle W_{a}\otimes W_{b}\rangle & = \sum_{i,\bar{i};k,\bar{k}}\epsilon_{i,\bar{i};k,\bar{k}}\\
\label{mumPCC2}
\langle W_{a}\otimes \mathbb{1}_{b}\rangle & = \sum_{i,\bar{i};k,k}\epsilon_{i,\bar{i};k,k}\\
\label{mumPCC3}
\langle\mathbb{1}_{a}\otimes  W_{b}\rangle & = \sum_{i,i;k,\bar{k}}\epsilon_{i,i;k,\bar{k}}
\end{align}
\subsubsection*{Proofs of Eqs.(\ref{numPCC1})-(\ref{mumPCC3})}
For 
$$\rho=\underset{i,j;k,l}{\sum} \epsilon_{i,j;k,l}|i\rangle\langle j|\otimes|k\rangle\langle l|$$
given by Eq.(\ref{eq:genbiparqudit})  we have

$$W_{a}\otimes W_{b}\rho=\underset{\bar{i},i,j;\bar{k},k,l}{\sum} \epsilon_{i,j;k,l}|\bar{i}\rangle\langle j|\otimes|\bar{k}\rangle\langle l|$$.\\

Note that, $\bar{i}\neq i$ and $\bar{k} \neq k$ by definition. Consequently, for the expectation value, we have
$$\langle W_{a}\otimes W_{b}\rho\rangle=\underset{i,\bar{i};k,\bar{k}}{\sum} \epsilon_{i,\bar{i};k,\bar{k}}$$

thus proving Eq.(\ref{numPCC1}).\\

Similarly, we can prove Eqs.(\ref{mumPCC2}) and (\ref{mumPCC3}).Then, combining Eqs.(\ref{numPCC1})-(\ref{mumPCC3}), we can obtain the following expression for the numerator of the expression for $PCC_{W}$ as follows: 
\begin{align}
\label{eq:numPCC}
\sum_{i,\bar{i};k,\bar{k}}\epsilon_{i,\bar{i};k,\bar{k}}-\left(\sum_{i,\bar{i};k}\epsilon_{i,\bar{i};k,k}\right)\left(\sum_{i;k,\bar{k}}\epsilon_{i,i;k,\bar{k}}\right)
\end{align}

Similarly, from Eq.(\ref{eq:genbiparqudit}) using $Z=\sum_{k}e_{k}|k\rangle\langle k|$, we can obtain the expression of the numerator of $PCC_{Z}$ as follows
\begin{align}
\label{eq:PCCZ}
 \sum_{j,k}e_{j}e_{k}\epsilon_{j,j;k,k} -\left(\sum_{j,k}e_{j}\epsilon_{j,j;k,k}\right)\left(\sum_{j,k}e_{k}\epsilon_{j,j;k,k}\right) 
\end{align}

Note that $Tr(W)=0$ as well as $Tr(Z)=0$ in the computational basis. Using Eqs.(\ref{eq:PCCZ}) and (\ref{eq:numPCC}) one can then seek to obtain the sum of PCCs for the various types of states and relate them to the Negativity for the respective states.\\

In what follows, we will show that for both pure and colored noise A states, the following relation holds good
\begin{equation}
\label{macgen}
PCC_{Z}+PCC_{W}-1 \propto \mathcal{N}
\end{equation}

\subsection{The sum of Pearson correlators for the pure and colored noise A states}

\subsubsection*{Pure state}
 
For pure entangled states we have the Schmidt bases $\lbrace |i\rangle \rbrace$ and $\lbrace |\alpha_{i}\rangle \rbrace$ such that
\begin{align}
\label{e1}
\ket{\psi}_{d} & = \sum_{i}\sqrt{\lambda_{i}}\ket{i}\ket{\alpha_{i}}\\
\label{e2}
\rho_{d} & = \sum_{i;j}\sqrt{\lambda_{i}\lambda_{j}}\ket{i}\bra{\alpha_{j}}\otimes\ket{i}\bra{\alpha_{j}}.
\end{align}

The Negativity for a pure state is given by
\begin{equation}
\label{eq:pure}
\mathcal{N}_{p}=\frac{1}{2}\sum_{i,j:i\neq j}\sqrt{\lambda_{i}\lambda_{j}}=\frac{1}{2}\sum_{i,\bar{i}}\sqrt{\lambda_{i}\lambda_{\bar{i}}}.
\end{equation}

Writing Eq.~(\ref{e2}) in the form of Eq.~(\ref{eq:genbiparqudit}) we obtain
\begin{equation}
\label{schmidtpure}
\epsilon_{i,j;k,l}=
\begin{cases}
\sqrt{\lambda_{i}\lambda_{j}} &  \text{for $i=k$ and $j=l$, $\epsilon_{i,j;i,j}$,}\\
$0$ & \text{otherwise.} 
\end{cases}
\end{equation}

Eq.~(\ref{schmidtpure}) implies that for $j=\bar{i}$ the non-zero terms are $\epsilon_{i,\bar{i};i,\bar{i}}=\sqrt{\lambda_{i}\lambda_{\bar{i}}}$. From Eq.~(\ref{numPCC1}) we obtain $\langle W_{a}\otimes W_{b}\rangle=\sum_{i,\bar{i}}\epsilon_{i,\bar{i};i,\bar{i}}=\sum_{i,\bar{i}}\sqrt{\lambda_{i}\lambda_{\bar{i}}}=2\mathcal{N}_{p}$. From Eqs.~(\ref{mumPCC2}) and (\ref{mumPCC3}) we have $\langle W_{a}\otimes\mathbb{1}\rangle=\langle \mathbb{1}\otimes W_{b}\rangle=0$. Consequently, the following relationship is obtained
\begin{equation}
    \label{preW1}
    PCC_{W}=\frac{2\mathcal{N}_{p}}{\Delta_{a}\Delta_{b}}
\end{equation}
 Here the variances $\Delta_{a},\Delta_{b}$ are given by $$\Delta_{a}=\sqrt{\langle W_{a}^{2}\otimes\mathbb{1}\rangle-\langle W_{a}\otimes\mathbb{1}\rangle}$$
$$\Delta_{b}=\sqrt{\langle \mathbb{1}\otimes W_{b}^{2}\rangle-\langle \mathbb{1}\otimes W_{b}\rangle}$$. 

Using Eq.~(\ref{sqhwo}) and the reasult that $\langle W_{a}\otimes\mathbb{1}\rangle=\langle \mathbb{1}\otimes W_{b}\rangle=0$ we can obtain 
$$\Delta_{a}=\Delta_{b}=\sqrt{d-1}$$
Thus, Eq.~(\ref{preW1}) can be rewritten as
\begin{equation}
    \label{preW}
    PCC_{W}=\frac{2\mathcal{N}_{p}}{(d-1)}
\end{equation}

Now, considering $PCC_{Z}$, note that since with respect to the Schmidt decomposition, the outcomes are perfectly correlated, it follows that
\begin{equation} 
    \label{preZ}
    PCC_{Z}=1
\end{equation}
Combining Eqs.(\ref{preW}) and (\ref{preZ}) we then obtain
\begin{equation}
    \label{pureMAC}
    PCC_{Z}+PCC_{W} = 1 + \frac{2\mathcal{N}_{p}}{(d-1)}
\end{equation}
Note that for the pure product state, the above sum is $1$. 

\subsubsection*{Colored noise A state}

First, we consider the colored noise A state defined as follows

\begin{align}
   \label{eq:CNA}
 \rho_{cna} & = \frac{p}{d}\sum_{j=0,k=0}^{d-1}\ket{j}\bra{k}\otimes\ket{j}\bra{k}+\frac{(1-p)}{d}\sum_{j=0}^{d-1}\ket{j}\bra{j}\otimes\ket{j}\bra{j}\\
 & = \frac{1}{d}\sum_{j}\ket{j}\bra{j}\otimes\ket{j}\bra{j}+\frac{p}{d}\sum_{j,\bar{j}}\ket{j}\bra{\bar{j}}\otimes\ket{j}\bra{\bar{j}} \nonumber
 \end{align}
 
 The Negativity $(\mathcal{N}_{a})$ of a colored noise A state is given by
 \begin{equation}
     \label{eq:neg_cna}
     \mathcal{N}_{a}=\frac{p(d-1)}{2}
 \end{equation}

Writing Eq.~(\ref{eq:CNA}) in the form of Eq.~(\ref{eq:genbiparqudit}) we obtain 
 \begin{equation}
\label{cnacompu}
    \epsilon_{i,j;k,l}=
    \begin{cases}
    \frac{1}{d} & \text{if $i=j=k=l$, $\epsilon_{j,j;j,j}$} \\
    \frac{p}{d} &  \text{if $i=k\neq j=l$, $\epsilon_{j,\bar{j};j,\bar{j}}$}\\
    \text{zero} & \text{otherwise}
    \end{cases}
\end{equation}

 It is evident from Eq.~(\ref{cnacompu}) that the terms $\epsilon_{j,j;k,\bar{k}}=\epsilon_{j,\bar{j};k,k}=0$. Consequently, R.H.S of Eqs.~(\ref{mumPCC2}) and (\ref{mumPCC3}) are zero and we have $\underset{j,\bar{j},k,\bar{k}}{\sum}\epsilon_{j,\bar{j};k,\bar{k}}=p(d-1)$.\\
 
 Using Eq.~(\ref{numPCC1}), we then have $\langle W_{a}\otimes W_{a}\rangle=p(d-1)=2\mathcal{N}_{a}$, whence the numerator of the expression of $PCC_{W}$ becomes $2\mathcal{N}_{a}$. Using Eq.(\ref{eq:sqW}), the denominator of $PCC_{W}$ is obtained as $(d-1)$. Thus, for the colored noise A we obtain 
 \begin{equation}
     \label{eq:cnawpcc}
     PCC_{W} = \frac{2\mathcal{N}_{a}}{(d-1)}
 \end{equation}
 
 Next, in order to obtain the numerator of the expression for $PCC_{Z}$ for the colored noise A state, we can use Eq.~(\ref{eq:PCCZ}) along with Eq.~(\ref{cnacompu}) which yield
 \begin{equation}
    \label{numcnaz}
    \langle Z_{a}\otimes Z_{b}\rangle-\langle Z_{a}\otimes \mathbb{1}_{b}\rangle    \langle\mathbb{1}_{a}\otimes  Z_{b}\rangle = \sum_{j}e_{j}^{2}.
 \end{equation}
 
 The denominator of the expression for $PCC_{Z}$ can then be obtained as follows
 
\begin{align}
\langle Z_{a}^{2}\otimes\mathbb{1}\rangle & = p\sum_{j}e_{j}^{2}+(1-p)\sum_{j}e_{j}^{2}\nonumber\\
\label{eq:var1}
& = \sum_{j}e_{j}^{2} = \langle \mathbb{1}\otimes Z_{b}^{2}\rangle
\end{align}
Using $Tr(Z_{a})=Tr(Z_{b})=\sum_{j}j=0$, one can verify that 
\begin{align}
\langle Z_{a}\otimes\mathbb{1}\rangle & = \langle \mathbb{1}\otimes Z_{b}\rangle\nonumber\\
& = \sum_{j}e_{j}\nonumber\\
\label{eq:var2}
& = 0
\end{align}
Combining Eqs.~(\ref{numcnaz})-(\ref{eq:var2}), the value of $PCC_{Z}$ is obtained as 
\begin{equation}
\label{eq:pcczzcna}
PCC_{Z}=1
\end{equation}
 Combining Eqs.~(\ref{eq:pcczzcna}) and (\ref{eq:cnawpcc}) the final relation is then derived as
 \begin{equation}
\label{macconjcna}
PCC_{Z}+PCC_{W}=1+\frac{2\mathcal{N}_{a}}{(d-1)}
\end{equation}

 To sum up, the sums of the two PCCs, $PCC_{Z}$ and $PCC_{W}$, as functions of the Negativity of the states considered given by the above derived Eqs.~(\ref{pureMAC}) and (\ref{macconjcna}) justify the conjecture made by Maccone et al.\cite{Maccone2015} for the bipartite arbitrary dimensional pure and colored noise A states respectively.  

\end{document}